\documentclass[rmp,amssymb,amsfonts,twocolumn]{revtex4}

\def\beq{\begin{equation}}
\def\eeq{\end{equation}}
\def\0{\otimes}
\def\1{\mbox{1\hskip-.25em l}}  
\def\6{\langle}
\def\9{\rangle}
\def\half{\mbox{$1\over2$}}
\def\pp{{\bf p}}
\def\tr{{\rm tr}\,}
\def\Eq{Eq.~(\ref}

\def\etal{{\it et al.\/}}
\def\cC{$\cal C$}
\def\cA{{\cal A}}
\def\cB{{\cal B}}
\def\cO{{\cal O}}
\def\kB{k_{{}_{{\rm B}}}} 
\def\BH{_{{}_{{\rm BH}}}} 
\def\lP{l_{{}_{{\rm P}}}} 

\def\bk{{\bf k}}

\def\bep{\mbox{\boldmath $\epsilon$}}
\def\bal{\mbox{\boldmath $\alpha$}}

\def\bb{{\bf b}}
\def\be{{\bf e}}
\def\bp{{\bf p}}
\def\bq{{\bf q}}
\def\bx{{\bf x}}
\def\by{{\bf y}}
\def\bz{{\bf z}}
\def\hbk{\hat{\bk}}
\def\hbx{\hat{\bx}}
\def\hby{\hat{\by}}
\def\hbz{\hat{\bz}}

\def\bay{\begin{array}}
\def\eay{\end{array}}

\def\Sc{Schwarzschild}

\def\cO{{\cal O}}

\def\half{\mbox{$1\over2$}}
\def\s12{\mbox{$1\over\sqrt{2}$}}
\usepackage{graphicx}
\usepackage{epsfig}

\begin{document}

\title{Quantum Information and Relativity
Theory}

\author{Asher Peres}
\affiliation{Department of Physics, Technion --- Israel Institute of
Technology, 32000 Haifa, Israel}
\author{and \\~  \\ Daniel R. Terno}
\affiliation{Perimeter Institute for Theoretical Physics, Waterloo,
Ontario, Canada N2J 2W9}

\begin{abstract}
Quantum mechanics, information theory, and relativity theory are
the basic foundations of theoretical physics. The acquisition of
information from a quantum system is the interface of classical
and quantum physics. Essential tools for its description are Kraus
matrices and positive operator valued measures (POVMs). Special
relativity imposes severe restrictions on the transfer of
information between distant systems. Quantum entropy is not a
Lorentz covariant concept. Lorentz transformations of reduced
density matrices for entangled systems may not be completely
positive maps. Quantum field theory, which is necessary for a
consistent description of interactions, implies a fundamental
trade-off between detector reliability and localizability. General
relativity produces new, counter\-intuitive effects, in particular
when black holes (or more generally, event horizons) are involved.
Most of the current concepts in quantum information theory may
then require a reassessment.

\end{abstract}
\maketitle

\tableofcontents

\section{Three inseparable theories}

Quantum theory and relativity theory emerged at the beginning of
the twentieth century to give answers to unexplained issues in
physics: the black body spectrum, the structure of atoms and
nuclei, the electrodynamics of moving bodies. Many years later,
information theory was developed by Claude Shannon (1948) for
analyzing the efficiency of communication methods.  How do these
seemingly disparate disciplines affect each other? In this review,
we shall show that they are inseparably related.

\subsection{Relativity and information}

Common presentations of relativity theory employ fictitious
observers who send and receive signals. These ``observers'' should
not be thought of as human beings, but rather ordinary physical
emitters and detectors. Their role is to label and locate events
in spacetime. The speed of transmission of these signals is
bounded by $c$ --- the velocity of light --- because information
needs a {\it material carrier\/}, and the latter must obey the
laws of physics. Information is physical (Landauer, 1991).

However, the mere existence of an upper bound on the speed of
propagation of physical effects does not do justice to the
fundamentally new concepts that were introduced by Albert Einstein
(one could as well imagine communications limited by the speed of
sound, or that of the postal service). Einstein showed that
simultaneity had no absolute meaning, and that distant events
might have different time orderings when referred to observers in
relative motion.  Relativistic kinematics is all about information
transfer between observers in relative motion.

Classical information theory involves concepts such as the rates
of emission and detection of signals, and the noise power
spectrum. These variables have well defined relativistic
transformation properties, independent of the actual physical
implementation of the communication system. A detailed analysis by
Jarett and Cover (1981) showed that the transmission rates for
observers with relative velocity $v$ were altered by a factor
$(c+v)/(c-v)$, namely the {\it square\/} of the familiar Doppler
factor for frequencies of periodic phenomena. We shall later
derive the same factor from classical electromagnetic theory, see
Eq.~(\ref{doppler}) below. Physics has a remarkably rigid
theoretical structure: you cannot alter any part of it without
having to change everything (Weinberg, 1992).

\subsection{Quantum mechanics and information}

Einstein's theory elicited strong opposition when it was proposed,
but is generally accepted by now. On the other hand, the
revolution caused by quantum theory still produces uneasy feelings
among some physicists.\footnote{The theory of relativity did not
cause as much misunderstanding and controversy as quantum theory,
because people were careful to avoid using the same nomenclature
as in nonrelativistic physics. For example, elementary textbooks
on relativity theory distinguish ``rest mass'' from ``relativistic
mass'' (hard core relativists call them simply ``mass'' and
``energy'').} Standard texbooks on quantum mechanics tell you that
observable quantities are represented by Hermitian operators,
their possible values are the eigenvalues of these operators, and
that the probability of detecting eigenvalue $\lambda_n$,
corresponding to eigenvector $u_n$, is $|\6u_n|\psi\9|^2$, where
$\psi$ is the (pure) state of the quantum system that is observed.
With a bit more sophistication to include mixed states, the
probability can be written in a general way $\6u_n|\rho|u_n\9$.

This is nice and neat, but this does not describe what happens in
real life. Quantum phenomena do not occur in a Hilbert space; they
occur in a laboratory. If you visit a real laboratory, you will
never find there Hermitian operators. All you can see are emitters
(lasers, ion guns, synchrotrons, and the like) and appropriate
detectors. In the latter, the time required for the irreversible
act of amplification (the formation of a microscopic bubble in a
bubble chamber, or the initial stage of an electric discharge) is
extremely brief, typically of the order of an atomic radius
divided by the velocity of light. Once irreversibility has set in,
the rest of the amplification process is essentially classical. It
is noteworthy that the time and space needed for initiating the
irreversible processes are incomparably smaller than the
macroscopic resolution of the detecting equipment.\footnote{The
``irreversible act of amplification'' is part of the quantum
folklore, but it is not essential to physics. Amplification is
solely needed to facilitate the work of the experimenter.}

The experimenter controls the emission process and observes
detection events. The theorist's problem is to predict the
probability of response of this or that detector, for a given
emission procedure. It often happens that the preparation is
unknown to the experimenter, and then the theory can be used for
discriminating between different preparation hypotheses, once the
detection outcomes are known.

Quantum mechanics tells us that whatever comes from the emitter is
represented by a state $\rho$ (a positive
operator,\footnote{Positive operators are those having the
property that $\6\psi|\rho|\psi\9\ge0$ for any state $\psi$. These
operators are always Hermitian.}  usually normalized to unit
trace). Detectors are represented by positive operators $E_\mu$,
where $\mu$ is an arbitrary label which identifies the detector.
The probability that detector $\mu$ be excited is $\tr(\rho
E_\mu)$. A complete set of $E_\mu$, including the possibility of
no detection, sums up to the unit matrix and is called a {\it
positive operator valued measure\/} (POVM). The various $E_\mu$ do
not in general commute, and therefore a detection event does not
correspond to what is commonly called the ``measurement of an
observable.'' Still, the activation of a particular detector is a
macroscopic, objective phenomenon. There is no uncertainty as to
which detector actually clicked.

Many physicists, perhaps a majority, have an intuitive realistic
worldview and consider a quantum state as a physical entity. Its
value may not be known, but in principle the quantum state of a
physical system would be well defined. However, there is no
experimental evidence whatsoever to support this naive belief. On
the contrary, if this view is taken seriously, it may lead to
bizarre consequences, called ``quantum paradoxes.'' These
so-called paradoxes originate solely from an incorrect
interpretation of quantum theory. The latter is thoroughly
pragmatic and, when correctly used, never yields two contradictory
answers to a well posed question. It is only the misuse of quantum
concepts, guided by a pseudorealistic philosophy, that leads to
paradoxical results.

In this review we shall adhere to the view that $\rho$ is only a
mathematical expression which encodes {\it information\/} about
the potential results of our experimental interventions.   The
latter are commonly called ``measurements'' --- an unfortunate
terminology, which gives the impression that there exists in the
real world some unknown property that we are measuring. Even the
very existence of particles depends on the context of our
experiments. In a classic article, Mott (1929) wrote ``Until the
final interpretation is made, no mention should be made of the
\mbox{$\alpha$-ray} being a particle at all.'' Drell (1978)
provocatively asked ``When is a particle?'' In particular,
observers whose world lines are accelerated record different
numbers of particles, as will be explained in Sec.~V.D (Unruh,
1976; Wald, 1994).

\subsection{Relativity and quantum theory}

The theory of relativity deals with the geometric structure of a
four-dimensional spacetime. Quantum mechanics describes properties
of matter. Combining these two theoretical edifices is a difficult
proposition. For example, there is no way of defining a
relativistic proper time for a quantum system which is spread all
over space. A proper time can in principle be defined for a
massive {\it apparatus\/} (``observer'') whose Compton wavelength
is so small that its center of mass has classical coordinates and
follows a continuous world-line. However, when there is more than
one apparatus, there is no role for the private proper times that
might be attached to the observers' world-lines. Therefore a
physical situation involving several observers in relative motion
cannot be described by a wave function with a relativistic
transformation law (Aharonov and Albert, 1981; Peres, 1995, and
references therein). This should not be surprising because a wave
function is not a physical object. It is only a tool for computing
the probabilities of objective macroscopic events.

Einstein's principle of relativity asserts that there are no
privileged inertial frames. This does not imply the necessity or
even the possibility of using manifestly symmetric
four-dimensional notations. This is {\it not\/} a peculiarity of
relativistic quantum mechanics. Likewise in classical canonical
theories, time has a special role in the equations of motion.

The relativity principle is extraordinarily restrictive. For
example, in ordinary classical mechanics with a finite number of
degrees of freedom, the requirement that the canonical coordinates
{\bf q} have the meaning of positions, so that particle
trajectories ${\bf q}(t)$ transform like four-dimensional world
lines, implies that these lines consist of straight segments. Long
range interactions are forbidden; there can be only contact
interactions between point particles (Currie, Jordan, and
Sudarshan, 1963; Leutwyler, 1965). Nontrivial relativistic
dynamics requires an {\it infinite\/} number of degrees of freedom
which are labelled by the spacetime coordinates (this is called a
field theory).

Combining relativity and quantum theory is not only a difficult
technical question on how to formulate dynamical laws. The
ontologies of these theories are radically different. Classical
theory asserts that fields, velocities, etc., transform in a
definite way and that the equations of motion of particles and
fields behave covariantly. For example if the expression for the
Lorentz force is written $f_\mu=F_{\mu\nu}u^\nu$ in one frame, the
same expression is valid in any other frame. These symbols
($f_\mu$, etc.) have objective values. They represent entities
that really exist, according to the theory. On the other hand,
wave functions are not defined in spacetime, but in a
multidimensional Hilbert space. They do not transform covariantly
when there are interventions by external agents, as will be seen
in Sec. III. Only the classical parameters attached to each
intervention transform covariantly. Yet, in spite of the
non-covariance of $\rho$, the final results of the calculations
(the probabilities of specified sets of events) must be Lorentz
invariant.

As a simple example, consider our two observers, conventionally
called Alice and Bob,\footnote{Alice and Bob joined the quantum
information community after a distinguished service in classical
crypto\-graphy. For example, they appeared in the historic RSA
paper (Rivest, Shamir, and Adleman, 1978).} holding a pair of
spin-$1\over2$ particles in a singlet state. Alice measures
$\sigma_z$ and finds +1, say. This tells her what the state of
Bob's particle is, namely the probabilities that Bob would obtain
$\pm1$ if he measures (or has measured, or will measure)
{\boldmath $\sigma$} along any direction he chooses. This is
purely counter\-factual information: {\it nothing\/} changes at
Bob's location until he performs the experiment himself, or
receives a message from Alice telling him the result that she
found. In particular, no experiment performed by Bob can tell him
whether Alice has measured (or will measure) her half of the
singlet.

A seemingly paradoxical way of presenting these results is to ask
the following naive question: suppose that Alice finds that
$\sigma_z=1$ while Bob does nothing. When does the state of Bob's
particle, far away, become the one for which $\sigma_z=-1$ with
certainty? Though this question is meaningless, it may be given a
definite answer: Bob's particle state changes instantaneously. In
which Lorentz frame is this instantaneous?  In {\it any\/} frame!
Whatever frame is chosen for defining simultaneity, the
experimentally observable result is the same, as can be shown in a
formal way (Peres, 2000b). Einstein himself was puzzled by what
seemed to be the instantaneous transmission of quantum
information.  In his autobiography, he wrote the words
``telepathically'' and ``spook'' (Einstein, 1949).

Examples like the above one, taken from relativistic quantum
mechanics, manifestly have an informational nature.  We cannot
separate the three disciplines: relativity, quantum mechanics, and
information theory.

 \subsection{The meaning of probability}
In this review, we shall often invoke the notion of {\it
probability\/}. Quantum mechanics is fundamentally statistical
(Ballentine, 1970). In the laboratory, any experiment has to be
repeated many times in order to infer a law; in a theoretical
discussion, we may imagine an infinite number of replicas of our
gedanken\-experiment, so as to have a genuine statistical
ensemble. Yet, the validity of the statistical nature of quantum
theory is not restricted to situations where there are a large
number of similar systems. {\it Statistical predictions do apply
to single events.\/} When we are told that the probability of
precipitation tomorrow is 35\%, there is only one tomorrow. This
tells us that it may be advisable to carry an umbrella.
Probability theory is simply the quantitative formulation of how
to make rational decisions in the face of uncertainty (Fuchs and
Peres, 2000).  A lucid analysis of how probabilistic concepts are
incorporated into physical theories is given by Emch and Liu
(2002).

\subsection{The role of topology}
Physicists often tend to ignore the topological structure of the
concepts that they use, or turn to it only as a last resort.
Actually, there is a ``bewildering'' multitude of topologies (Reed
and Simon, 1980). Many of them have a direct physical meaning
(Emch 1972; Haag, 1996; Araki, 1999). In particular, since
measurements can actually be performed only with a finite
accuracy, a finite number of outcomes, and a finite number of
times, only bounded ranges of values are ever registered. Suppose
that we measure $N$ times the value $q$ of an observable $Q$, and
a value $q_j$ is obtained $n_j$ times. A relative frequency
$w_j=n_j/N$ is either used to extract a probability estimate, or
it is taken at face value and interpreted as the estimate. Thus
the information about a state $\rho$ can be formulated as (Araki,
1999; Peres and Terno, 1998)
\beq
|p_\rho^Q(q_j)-w_j|<\epsilon_j,
\eeq
for some positive $\epsilon_j$. These inequalities induce a
natural topology on the space of states, which is called a
``physical topology'' (Emch, 1972; Araki, 1999). More precisely,
they define a weak-* topology on the  observables and a weak
topology on the states. This is a trace-norm
topology\footnote{Since probabilities in quantum mechanics are
given by the expression $\tr(\rho E_\mu)$, and physically
acceptable states are trace class positive operators, the trace
norm topology is the concrete realization of the physical
topology.} (Reed and Simon, 1980). These structures are naturally
accommodated in the algebraic approach to quantum theory. That
approach consists in the characterization of the theory by a net
of algebras of local observables, and is especially suited for the
analysis of infinite systems in quantum statistical mechanics and
quantum field theory. We will use results based on algebraic field
theory in Sec.~V and VI.\footnote{References whose primary
interest is field theory include Bogolubov
\etal\/ (1990), Haag (1996) and Araki (1999). On the other hand, Davies
(1976), Bratteli and Robinson (1987), and Ingraden, Kossakowski
and Ohaya (1997), consider mainly applications to open quantum
systems, statistical mechanics and thermodynamics. Emch (1972) is
concerned with both. Emch (1972), Bratelli and Robinson (1987),
and Baumgartel and Wollenberg (1992) give a rigorous, and yet
readable exposition of the subject. }

\subsection{The essence of quantum information}
In an early review of quantum information theory, Ingarden (1976)
distinguished two fundamental aspects:
\begin{quote}
``Information theory, as it is understood in this paper and as it
usually understood by mathematicians and engineers following the
pioneer paper of Shannon, is not only a theory of the entropy
concept itself (in this aspect information theory is most
interesting for physicists), but also a theory of transmission and
coding of information, {\it i.e.\/}, a theory of information
sources and channels.''
\end{quote}
In other words: the goals of quantum information theory are the
intersection of those of quantum mechanics and information theory,
while its tools are the union of those of these two theories.
Actually, the tools belonging to quantum theory were developed
under the influence of nascent quantum information, ``when it was
necessary to consider communication problems for the needs of
quantum of quantum electronic and optics" (Ingarden, 1976). Work
of Sudarshan \etal\ (1961), and later those of Davies,
Kossakowski, Kraus,  Lindblad, and Lewis established the formalism
of quantum mechanics of open systems, expressed by POVMs and
completely positive maps, while works of Helstrom, Holevo,
Lebedev, and Levitin, produced important results in what became
quantum information theory.\footnote{The books of Davies (1976),
Holevo (1982), and  Ingarden, Kossakowski and Ohaya (1997),
contain historical surveys and exhaustive lists of references. }
We shall discuss these subjects in Sec.~II of this review.

Some trends in modern quantum information theory may be traced to
security problems in quantum communication. A very early
contribution was Wiesner's seminal paper {\it Conjugate Coding\/},
which was submitted circa 1970 to IEEE Transactions on Information
Theory, and promptly rejected because it was written in a jargon
incomprehensible to computer scientists (this actually was a paper
about physics, but it had been submitted to a computer science
journal). Wiesner's article was finally published (Wiesner, 1983)
in the newsletter of ACM SIGACT (Association for Computing
Machinery, Special Interest Group in Algorithms and Computation
Theory). That article tacitly assumed that exact duplication of an
unknown quantum state was impossible, well before the no-cloning
theorem (Wootters and Zurek, 1982; Dieks, 1982) became common
knowledge. Another early article, {\it Unforgeable Subway
Tokens\/} (Bennett \etal, 1983), also tacitly assumed the same.

The standard method for quantum cryptography was invented by
Bennett and Brassard (1984), using two {\it mutually unbiased
bases\/}, namely two bases such that $\6u_m|v_\mu\9=1/\sqrt{d}$,
where $d$ is the number of Hilbert space dimensions. Security may
be improved by using three bases (Bru\ss, 1998;
Bechmann-Pasquinucci and Gisin, 1999), and even more by going to
higher dimensions (Bechmann-Pasquinucci and Peres, 2000; Bru\ss\
and Macchiavello 2002). Gisin, Ribordy, Tittel and Zbinden (2002)
recently reviewed theoretical and experimental results in quantum
cryptography.

A spectacular discovery was that of quantum teleportation (Bennett
\etal., 1993), which effectively turned quantum entanglement
into a communication resource. Soon afterwards, it also became a
computational resource (Shor, 1994) and since then it continues to
attract considerable attention. Various aspects of entanglement
theory are reviewed in special issues of Quantum Information and
Computation (2001) {\bf 1} (1) and Journal of Mathematical Physics
(2002) {\bf43} (9). Experimental results were reviewed by
Zeilinger (1999).

Quantum binary channels were introduced by Schumacher (1995), who
also generalized Shannon's coding theorems to the quantum domain,
and coined the word ``qubit" (quantum bit) for elementary carriers
of quantum information. Quantum channels are discussed by Holevo
(1999), Amosov, Holevo, and Werner (2000), King and Ruskai (2001),
and in the special issue of Journal of Mathematical Physics (2002)
{\bf43} (9). An extensive review of the mathematical aspects of
quantum information theory was given by Keyl (2002).

Our review deals with many interrelated issues.  Causality
constraints on POVMs are discussed in Sec.~II.E.  Relativistic
extensions of the formalism appear in Sec.~III and VI.A. In
Sec.~IV we discuss how relativistic considerations modify basic
notions of quantum information theory: qubits, entanglement, and
quantum channels. In Sec.~V we investigate the implications of
quantum field theory on the construction of POVMs and the
detection of entanglement.  Section VI.A deals with relativistic
extensions of quantum information theory, and in Sec.~VI.B we
discuss its applications to the black hole physics.

\section{The acquisition of information}
\subsection{The ambivalent quantum observer}

Quantum mechanics is used by theorists in two different ways: it
is a tool for computing accurate relationships between physical
constants, such as energy levels, cross sections, transition
rates, etc. These calculations are technically difficult, but they
are not controversial.  Besides this, quantum mechanics also
provides statistical predictions for results of measurements
performed on physical systems that have been prepared in a
specified way. The quantum measuring process is the interface of
classical and quantum phenomena.  The preparation and measurement
are performed by macroscopic devices, and these are described in
classical terms. The necessity of using a classical terminology
was emphasized by Niels Bohr (1927) since the very early days of
quantum mechanics. Bohr's insistence on a classical description
was very strict. He wrote (1949):

\begin{quote}``\ldots by the word `experiment' we refer to a situation
where we can tell others what we have done and what we have
learned and that, therefore, the account of the experimental
arrangement and of the results of the observations must be
expressed in unambiguous language, with suitable application of
the terminology of classical physics.'' \end{quote}

Note the words ``we can tell.'' Bohr was concerned with {\it
information\/}, in the broadest sense of this term. He never said
that there were classical systems or quantum systems. There were
physical systems, for which it was appropriate to use the
classical language or the quantum language. There is no guarantee
that either language gives a perfect description, but in a well
designed experiment it should be at least a good approximation.

Bohr's approach divides the physical world into ``endosystems''
(Finkelstein, 1988) that are described by quantum dynamics, and
``exosystems'' (such as measuring apparatuses) that are not
described by the dynamical formalism of the endosystem under
consideration. A physical system is called ``open'' when parts of
the universe are excluded from its description. In different
Lorentz frames used by observers in relative motion, different
parts of the universe may be excluded. The systems considered by
these observers are then essentially different, and no Lorentz
transformation exists that can relate them (Peres and Terno,
2002).

It is noteworthy that Bohr never described the measuring process
as a dynamical interaction between an exophysical apparatus and
the system under observation. He was of course fully aware that
measuring apparatuses are made of the same kind of matter as
everything else, and they obey the same physical laws. It is
therefore tempting to use quantum theory in order to investigate
their behavior during a measurement. However, if this is done, the
quantized apparatus loses its status of a measuring instrument. It
becomes a mere intermediate system in the measuring process, and
there must still be a {\it final\/} instrument that has a purely
classical description (Bohr, 1939).

Measurement was understood by Bohr as a primitive notion. He could
thereby elude questions which caused considerable controversy
among other authors.  A quantum dynamical description of the
measuring process was first attempted by John von Neumann, in his
treatise on the mathematical foundations of quantum theory (1932).
In the last section of that book, as in an after\-thought, von
Neumann represented the apparatus by a single degree of freedom,
whose value was correlated to that of the dynamical variable being
measured.  Such an apparatus is not, in general, left in a
definite pure state, and it does not admit a classical
description. Therefore, von Neumann introduced a second apparatus
which observes the first one, and possibly a third apparatus, and
so on, until there is a final measurement, which is {\it not\/}
described by quantum dynamics and has a definite result (for which
quantum mechanics can only give statistical predictions). The
essential point that was suggested, but not proved by von Neumann,
is that the introduction of this sequence of apparatuses is
irrelevant: the final result is the same, irrespective of the
location of the ``cut'' between classical and quantum
physics.\footnote{At this point, von Neumann also speculated that
the final step involves the consciousness of the observer --- a
bizarre statement in a mathematically rigorous monograph (von
Neumann, 1955).}

These different approaches of Bohr and von Neumann were reconciled
by Hay and Peres (1998), who introduced a dual description for the
measuring apparatus. It obeys quantum mechanics while it interacts
with the  system under observation, and then it is ``dequantized''
and is described by a classical Liouville density which provides
the probability distribution for the results of the measurement.
Alternatively, the apparatus may always be treated by quantum
mechanics, and be measured by a second apparatus which has such a
dual description. The question raised by Hay and Peres is whether
these two different methods of calculation give the same result,
or at least asymptotically agree under suitable conditions.  They
showed that a sufficient condition for agreement between the two
methods is that the dynamical variable used as a ``pointer'' by
the first apparatus be represented by a ``quasi-classical''
operator of the Weyl-Wigner type (Hillery
\etal, 1984).

To avoid any misunderstanding, we emphasize that the classical
description of a pointer is not by means of a point in phase
space, but by a {\it Liouville density\/}. Quantum theory makes
only statistical predictions, and any semi\-classical treatment
that simulates it must also be statistical.

\subsection{The measuring process}

Dirac (1947) wrote ``a measurement always causes the system to
jump into an eigenstate of the dynamical variable being
measured.'' Here, we must be careful: a quantum jump (also called
{\it collapse\/}) is something that happens in our description of
the system, not to the system itself. Likewise, the time
dependence of the wave function does not represent the evolution
of a physical system.  It only gives the evolution of
probabilities for the outcomes of potential experiments on that
system (Fuchs and Peres, 2000).

Let us examine more closely the measuring process. First, we must
refine the notion of measurement and extend it to a more general
one: an {\it intervention\/}.  An intervention is described by a
set of parameters which include the location of the intervention
in spacetime, referred to an arbitrary coordinate system. We also
have to specify the speed and orientation of the apparatus in the
coordinate system that we are using, and various other input
parameters that control the apparatus, such as the strength of a
magnetic field, or that of an rf pulse used in the experiment. The
input parameters are determined by classical information received
from past interventions, or they may be chosen arbitrarily by the
observer who prepares that intervention, or by a local random
device acting in lieu of the observer.

An intervention has two consequences. One is the acquisition of
information by means of an apparatus that produces a record. This
is the ``measurement.'' Its outcome, which is in general
unpredictable, is the output of the intervention. The other
consequence is a change of the environment in which the quantum
system will evolve after completion of the intervention. For
example the intervening apparatus may generate a new Hamiltonian
which depends on the recorded result. In particular, classical
signals may be emitted for controlling the execution of further
interventions. These signals are of course limited to the velocity
of light.

The experimental protocols that we consider all start in the same
way, with the same initial state $\rho_0$, and the first
intervention is the same. However, later stages of the experiment
may involve different types of interventions, possibly with
different spacetime locations, depending on the outcomes of the
preceding events.  Yet, assuming that each intervention has only a
finite number of outcomes, there is for the entire experiment only
a finite number of possible records. (Here, the word ``record''
means the complete list of outcomes that occurred during the
experiment. We do not want to use the word ``history'' which has
acquired a different meaning in the writings of some quantum
theorists.)

Each one of these records has a definite probability in the
statistical ensemble. In the laboratory, experimenters can observe
its relative frequency among all the records that were obtained;
when the number of records tends to infinity, this relative
frequency is expected to tend to the true probability. The aim of
theory is to predict the probability of each record, given the
inputs of the various interventions (both the inputs that are
actually controlled by the local experimenter and those determined
by the outputs of earlier interventions). Each record is
objective: everyone agrees on what happened (e.g., which detectors
clicked). Therefore, everyone agrees on what the various relative
frequencies are, and the theoretical probabilities are also the
same for everyone.

Interventions are localized in spacetime, but quantum systems are
pervasive. In each experiment, irrespective of its history, there
is only one quantum system, which may consist of several particles
or other subsystems, created or annihilated at the various
interventions. Note that all these properties still hold if the
measurement outcome is the {\it absence\/} of a detector click. It
does not matter whether this is due to an imperfection of the
detector or to a probability $<1$ that a perfect detector would be
excited. The state of the quantum system does not remain
unchanged.  It has to change to respect unitarity. The mere
presence of a detector that could have been excited implies that
there has been an interaction between that detector and the
quantum system. Even if the detector has a finite probability of
remaining in its initial state, the quantum system correlated to
the latter acquires a different state (Dicke, 1981). The absence
of a click, when there could have been one, is also an event.

Interventions, as defined above, start by an interaction with a
measuring apparatus, called ``premeasurement'' (Peres, 1980). The
quantum system and the apparatus are initially in a state $\sum_s
c_s\,|s\9\0|A\9$, and become entangled into a single composite
system \cC:
\beq
\sum_s c_s\,|s\9\0|A\9\to
  \sum_{s,\lambda} c_s\,U_{s\lambda}\,|\lambda\9,\label{premea}
\eeq
where $\{|\lambda\9\}$ is a complete basis for the states of \cC.
It is the choice of the unitary matrix $U_{s\lambda}$ that
determines which property of the system under study is correlated
to the apparatus, and therefore is measured. When writing the
above equation, we tacitly assumed that the quantum system and the
measuring apparatus were initially in a pure state. Since a mixed
state is a convex combination of pure states, no new feature can
result from taking mixed states (which would admittedly be more
realistic). Relativistic restrictions on the allowed forms of
$U_{s\lambda}$ will be discussed in Sec.~III.

The measuring process involves not only the physical system under
study and a measuring apparatus (which together form the composite
system
\cC) but also their ``environment'' which includes unspecified
degrees of freedom of the apparatus and the rest of the world.
These unknown degrees of freedom interact with the relevant ones,
but they are not under the control of the experimenter and cannot
be explicitly described. Our partial ignorance is not a sign of
weakness. It is fundamental. If everything were known, acquisition
of information would be a meaningless concept.

A complete description of \cC\ involves both macroscopic and
microscopic variables. The difference between them is that the
environment can be considered as adequately isolated from the
microscopic degrees of freedom for the duration of the experiment
and is not influenced by them, while {\it the environment is not
isolated from the macroscopic degrees of freedom\/}. For example,
if there is a macroscopic pointer, air molecules bounce from it in
a way that depends on the position of that pointer.  Even if we
can neglect the Brownian motion of a massive pointer, its
influence on the environment leads to the phenomenon of
decoherence, which is inherent to the measuring process.

An essential property of the composite system \cC, which is
necessary to produce a meaningful measurement, is that its states
form a finite number of orthogonal subspaces which are
distinguishable by the observer. Each macroscopically
distinguishable subspace corresponds to one of the outcomes of the
intervention and defines a POVM element $E_\mu$, given explicitly
by Eq.~(\ref{Emu}) below.  Let us therefore introduce a complete
basis for \cC, namely $\{|\mu,\xi\9\}$, where $\mu$ labels a
macroscopic subspace, and $\xi$ labels microscopic states in that
subspace.

\subsection{Decoherence}

Up to now, the quantum evolution is well defined and it is in
principle reversible. It would remain so if the environment could
be perfectly isolated from the macroscopic degrees of freedom of
the apparatus. This demand is of course self-contradictory, since
we have to read the result of the measurement if we wish to make
any use of it. A detailed analysis of the interaction with the
environment, together with plausible hypotheses (Peres, 2000a),
shows that states of the environment that are correlated to
subspaces of \cC\ with different labels $\mu$ can be treated as if
they were orthogonal. This is an excellent {\it approximation\/}
(physics is not an exact science, it is a science of
approximations).  The resulting theoretical predictions will
almost always be correct, and if any rare small deviation from
them is ever observed, it will be considered as a statistical
quirk, or an experimental error.

The density matrix of the quantum system thus is effectively
block-diagonal and all our statistical predictions are identical
to those obtained for an ordinary mixture of (unnormalized) pure
states
\beq
|\psi_\mu\9=\sum_{s,\xi}c_s\,U_{s\mu\xi}\,|\mu,\xi\9,
 \label{psimu}
 \eeq
where the statistical weight of each state is the square of its
norm. This process is called decoherence. Each subspace $\mu$ is
stable under decoherence --- it is their relative phase that
decoheres. From this moment on, the macroscopic degrees of freedom
of \cC\ have entered into the classical domain. We can safely
observe them and ``lay on them our grubby hands'' (Caves, 1982).
In particular, they can be used to trigger amplification
mechanisms (the so-called detector clicks) for the convenience of
the experimenter.

Some authors claim that decoherence may provide a solution of the
``measurement problem,'' with the particular meaning that they
attribute to that problem (Zurek, 1991). Others dispute this point
of view in their comments on the above article (Zurek, 1993). A
reassessment of this issue and many important technical details
were recently published by Zurek (2002, 2003). Yet, decoherence
has an essential role, as explained above. It is essential to
distinguish decoherence, which results from the disturbance of the
environment by the apparatus (and is a quantum effect), from {\it
noise\/}, which would result from the disturbance of the system or
the apparatus by the environment and would cause errors. Noise is
a mundane classical phenomenon, which we ignore in this
review.\footnote{The so-called ``quantum noise'' which is
discussed in Sec.~IV.C has a different nature.}

\subsection{Kraus matrices and POVMs}

The final step of the intervention is to discard part of the
composite system \cC. The discarded part may depend on the outcome
$\mu$. We therefore introduce in the subspace $\mu$ two sets of
basis vectors $|\mu,\sigma\9$ and $|\mu,m\9$ for the new system
and the part that is discarded, respectively. We thus obtain for
the new system a reduced density matrix
\beq
(\rho'_\mu)_{\sigma\tau}=\sum_{m}\sum_{s,t}
 (A_{\mu m})_{\sigma s}\;\rho_{st}\;(A_{\mu m}^*)_{\tau t},
 \label{rhomu}
 \eeq
where $\rho_{st}\equiv c_sc_t^*$ is the initial state, and the
notation
\beq
(A_{\mu m})_{\sigma s}\equiv U_{s\mu\sigma m}, \label{AU}
\eeq
was introduced for later convenience. Recall that the indices $s$
and $\sigma$ refer to the original system under study and to the
final one, respectively. Omitting these indices, \Eq{rhomu}) takes
the familiar form
\beq
\rho\to\rho'_\mu=\sum_m A_{\mu m}\,\rho\,A_{\mu m}^\dagger,
  \label{ArhoA}
\eeq
where $\mu$ is a label that indicates which detector was involved
and the label $m$ refers to any subsystem that was discarded at
the conclusion of the interaction. Clearly, the ``quantum jump''
$\rho\to\rho'_\mu$ is not a dynamical process that occurs in the
quantum system by itself. It results from the introduction of an
apparatus, followed by its deletion or that of another subsystem.
A jump in the quantum state occurs even when there is no detector
click or other macroscopic amplification, because we impose abrupt
changes in our way of delimiting the object that we consider as
the quantum system under study.

The initial $\rho$ is usually assumed to be normalized to unit
trace, and the trace of $\rho'_\mu$ is the probability of
occurrence of outcome $\mu$. Note that each symbol $A_{\mu m}$  in
the above equation represents a {\it matrix\/} (not a matrix
element).  Explicitly, the Kraus operators $A_{\mu m}$ (Kraus,
1983) are given by Eq.~(\ref{AU}), where $U_{s\mu\sigma m}$ is the
matrix element for the unitary interaction between the system
under study and the apparatus, including any auxiliary systems
that are subsequently discarded (Peres, 2000a).

Equation (\ref{ArhoA}) is sometimes written $\rho'_\mu={\cal
S}\rho$, where $\cal S$ is a linear {\it super\-operator\/} which
acts on density matrices like ordinary operators act on pure
states. Note however that these super\-operators have a very
special structure, explicitly given by Eq.~(\ref{ArhoA}).

It is noteworthy that Eq.~(\ref{ArhoA}) is the most general
completely positive linear map (Stinespring, 1955; Davies, 1976;
Kraus, 1983). This is a crucial property: a linear map $T(\rho)$
is called {\it positive\/} if it transforms any positive matrix
$\rho$ (namely, one without negative eigenvalues) into another
positive matrix. It is called {\it completely positive\/} if
$(T\otimes\1)$ acting on a bipartite $\rho$ produces a valid
bipartite $\rho$. For instance, complex conjugation of $\rho$
(whose meaning is time reversal) is a positive map. However, it is
not completely positive. If we have two systems, it is physically
meaningless to reverse the direction of time for only one of them.
One can write a formal expression for this impossible process, but
the resulting ``density matrix'' is unphysical because it may have
negative eigenvalues (Peres, 1996).  The case for consideration of
completely positive maps was made by Kraus (1971), Davies (1976)
and Lindblad (1976), and since than they are part of the
 toolbox of quantum information. In Sec.~\ref{cocha} we discuss apparent exceptions to
this approach.

It follows from Eq.~(\ref{ArhoA}) that the probability of
occurrence of outcome $\mu$ is
\beq
p_\mu=\sum_m\tr(A_{\mu m}\,\rho\,A_{\mu m}^\dagger)=
  \tr(\rho E_\mu).
\eeq
The positive operators
\beq
E_\mu=\sum_m A_{\mu m}^\dagger\,A_{\mu m}, \label{Emu}
\eeq
whose dimensions are the same as those of the initial $\rho$,
satisfy $\sum_\mu E_\mu=\1$ owing to the unitarity of
$U_{s\mu\sigma m}$. Therefore they are the elements of a POVM.
Conversely, given $E_\mu$ (a positive matrix of order $k$) it is
always possible to split it in infinitely many ways as in the
above equation.

In the special case where the POVM elements $E_\mu$ commute, they
are orthogonal projection operators, and the POVM becomes a {\it
projection valued measure\/} (PVM). The corresponding intervention
is sometimes called a von~Neumann measurement.  Rigorous treatment
of the POVM formalism can be found in the books of Davies (1976),
Holevo (1982), and Kraus (1983).

\subsection{The no-communication theorem}\label{no-com}

We now derive a sufficient condition that no instantaneous
information transfer can result from a distant intervention. We
shall show that the condition is
\beq
[A_{\mu m}, B_{\nu n}]=0, \label{etcr}
\eeq
where $A_{\mu m}$ and $B_{\nu n}$ are Kraus matrices for the
observation of outcomes $\mu$ by Alice and $\nu$ by Bob.  Indeed,
the probability that Bob gets a result $\nu$, irrespective of what
Alice found, is
\beq p_\nu=\sum_\mu\tr\Bigl(\,\sum_{m,n}B_{\nu n}\,A_{\mu m}\,\rho\,
  A^\dagger_{\mu m}\,B^\dagger_{\nu n}\Bigr). \label{pnu}
\eeq

We now make use of \Eq{etcr}) to exchange the positions of $A_{\mu
m}$ and $B_{\nu n}$, and likewise those of $A^\dagger_{\mu m}$ and
$B^\dagger_{\nu n}$, and then we move $A_{\mu m}$ from the first
position to the last one in the product of operators in the traced
parenthesis. We thereby obtain expressions as in Eq.~(\ref{Emu}).
These are elements of a POVM that satisfy $\sum_\mu E_\mu=\1$.
Therefore \Eq{pnu}) reduces to
\beq
p_\nu=\tr\Bigl(\sum_n B_{\nu n}\,\rho\,B^\dagger_{\nu n}\Bigr),
\eeq
whence all expressions involving Alice's operators $A_{\mu m}$
have totally disappeared. The statistics of Bob's result are not
affected at all by what Alice may simultaneously do somewhere
else. This proves that Eq.~(\ref{etcr}) indeed is a sufficient
condition for no instantaneous information transfer.\footnote{ An
algebraic approach to statistical independence and to related
topics is discussed by Florig and Summers (1997), while Neumann
and Werner (1983) specifically address the issue of causality
between preparation and registration processes.}

Note that any classical communication between distant observers
can be considered as a kind of long range interaction. Indeed, it
is always possible to treat their apparatuses as quantum systems
(von Neumann, 1932; Bohr, 1939) and then any signals that
propagate between these apparatuses are a manifestation of their
mutual interaction. The propagation of signals is of course
bounded by the velocity of light. As a result, there exists a
partial time ordering of the various interventions in an
experiment, which defines the notions earlier and later (we assume
that there are no closed causal loops). The input parameters of an
intervention are deterministic (or possibly stochastic) functions
of the parameters of earlier interventions, but not of the
stochastic outcomes resulting from later or mutually spacelike
interventions (Blanchard and Jadczik, 1996 and 1998; Percival,
1998).

Even these apparently simple notions lead to non-trivial results.
Consider a separable bipartite superoperator $T$,
\beq
T(\rho)=\sum_k M_k\rho M_k^\dag, \qquad M_k=A_k\otimes B_k,
\eeq
where the operators $A_k$ represent operations of Alice, and $B_k$
those of Bob. It was shown by Bennett \etal\ (1999) that not all
such superoperators can be implemented by local transformations
and classical communication (LOCC). For more on this subject, see
Walgate and Hardy (2002).

A classification of bipartite state transformations was introduced
by Beckman \etal\ (2001). It consists of the following categories.
There are {\it localizable\/} operations that can be implemented
locally by Alice and Bob, possibly with the help of prearranged
entangled auxiliary systems (ancillas), but without classical
comunication. Ideally, local operations are instantaneous, and the
whole process can be viewed as performed at a definite time. For
{\it semilocalizable\/} operations, the requirement of no
communication is relaxed and one-way classical communication is
possible. It is obvious that any tensor-product operation $T_{\rm
A}\otimes T_{\rm B}$ is localizable. The converse is not always
true, for example in Bell measurements (Braunstein, Mann, and
Revzen, 1992) which distinguish between the four standard
bipartite entangled states,
\beq |\Psi^\pm\9 := {1\over \sqrt 2}(|0\9 |1\9 \pm
|1\9 |0\9), \label{psipm}\eeq
\beq |\Phi^\pm\9 := {1\over \sqrt
2}(|0\9 |0\9 \pm |1\9 |1\9).\label{phipm}
\eeq

Other classes of bipartite operations are defined as follows: Bob
performs a local operation $T_{\rm B}$ just before the global
operation $T$. If no local operation of Alice can reveal any
information about $T_{\rm B}$, {\it i.e.\/}, Bob cannot signal to
Alice, then the operation $T$ is {\it semicausal\/}. If the
operation is semicausal in both directions, it is called {\it
causal\/}.

In many cases it is easier to prove causality than localizability.
To check the causality of an operation $T$ whose outcomes are the
states $\rho_\mu=T_\mu(\rho)/p_\mu$ with probabilities $p_\mu=\tr
T_\mu(\rho)$, it is enough to consider the corresponding
superoperator
\beq
T'(\rho):=\sum_\mu T_\mu(\rho).
\eeq
Indeed, assume that Bob's action prior to the global operation
leads to one of the two different states $\rho_1$ and $\rho_2$.
Then the states $T'(\rho_1)$ and $T'(\rho_2)$ are distinguishable
if and only if some of the pairs of states
$T_\mu(\rho_1)/p_{\mu1}$ and $T_\mu(\rho_2)/p_{\mu2}$ are
distinguishable. Such probabilistic distinguishability shows that
the operation $T$ is not semicausal. These definitions of causal
and localizable operators appear equivalent. It is easily proved
that localizable operators are causal. It was shown that
semicausal operators are always semilocalizable (Eggeling,
Schlingemann, and Werner, 2002). However, there are causal
operations that are not localizable (Beckman \etal, 2001).

It is curious that while a complete Bell measurement is causal,
the two-outcome incomplete Bell measurement is not (Sorkin, 1993).
Indeed, consider a two-outcome PVM
\beq
E_1=|\Phi^+\9\6\Phi^+|, \qquad E_2=\1-E_1,
\eeq
where $|\Phi^+\9=(|00\9+|11\9)/\sqrt{2}$ (and the Kraus matrices
are the projectors $E_\mu$ themselves). If the initial state is
$|01\9_{\rm AB}$, then the outcome that is associated with $E_2$
always occurs and Alice's reduced density matrix after the
measurement is $\rho_{\rm A}=|0\9\6 0|$. On the other hand, if
before the joint measurement Bob performs a unitary operation that
transforms the state into $|00\9_{\rm AB}$, then the two outcomes
are equiprobable, the resulting states after the measurement are
maximally entangled, and Alice's reduced density matrix is
$\rho_{\rm A}=\half\1$. It can be shown that two input states
$|00\9_{\rm AB}$ and $|01\9_{\rm AB}$ after this incomplete Bell
measurement are distinguished by Alice with a probability of 0.75.

Here is another example of a  semicausal and semilocalizable
measurement which can be executed with one-way classical
communication from Alice to Bob.  Consider a PVM measurement,
whose complete orthogonal projectors are
\beq
|0\9\0|0\9, \quad |0\9\0|1\9, \quad |1\9\0|+\9, \quad |1\9\0|-\9,
\label{examp1}
\eeq
where $|\pm\9=(|0\9\pm|1\9)/\sqrt{2}$. The Kraus matrices are
\beq
A_{\mu j}=E_\mu\delta_{j0}\label{orthopvm}.
\eeq
{}From the properties of complete orthogonal measurements (Beckman
\etal, 2001), it follows that this operation cannot be performed
without Alice talking to Bob. A protocol to realize this
measurement is the following. Alice measures her qubit in the
basis $\{|0\9, |1\9\}$, and tells her result to Bob. If Alice's
outcome was $|0\9$, Bob measures his qubit in the basis $\{|0\9,
|1\9\}$, and if it was $|1\9$, in the basis $\{|+\9, |-\9\}$.

Beckman \etal\ (2001) derived necessary and sufficient conditions
to check the semicausality (and therefore, the causality) of PVM
measurements. Groisman and Reznik (2002) allowed for more
complicated conditional state evolutions. In particular, they were
interested in {\em verification\/} measurements, {\it i.e.\/},
those yielding $\mu$ with certainty if the state prior to the
classical intervention is $\rho\propto E_\mu$, but without making
any specific demand on the resulting state $\rho'_\mu$. They
showed that all PVM verification measurements on $2\times 2$
dimensional systems are localizable.

Vaidman (2003) proposed a realization of verification measurements
by means of a shared entangled ancilla, and Bell-type measurements
by one of the parties. A verification measurement of the states in
Eq.~(\ref{examp1}) will illustrate his construction. Alice and Bob
share a Bell state $|\Psi^-\9$ and, contrary to the scheme of
Beckman
\etal\ (2001), they do not have  to coordinate their moves. Alice
and Bob perform their tasks independently and convey their results
to a common center, where the final analysis is made. In the first
step of this measurement, Alice performs a Bell measurement as in
the teleportation of a state $|\Psi\9$ from her site to Bob (see
below). However, Alice and Bob do not perform the full
teleportation which requires a classical communication between
them. The second step of the verification is executed by Bob. He
measures the spin of his particle in the $z$ direction. According
to whether that spin is up or down, he measures the spin of his
ancilla in the $z$ or $x$ direction, respectively. This completes
the measurement and it only remains to combine the local outcomes
to get the result of the nonlocal measurement (Vaidman, 2003).
This method can be extended to arbitrary Hilbert space dimensions.

In the teleportation of an unknown state $|\Psi\9_0$ of a
spin-\half\ particle located at Alice's site, Alice and Bob use a
prearranged pair in a singlet state, namely $|\Psi^-\9_{12}=
(|0\9_1 |1\9_2 - |1\9_1 |0\9_2)/\sqrt 2$. The procedure is based
on the identity (Bennett \etal, 1993)
\begin{widetext}
\beq
    |\Psi\9_0|\Psi^-\9_{12} = \half
  \left(|\Psi^-\9_{01} | \Psi\9_{2} +
|\Psi^+\9_{01} |\tilde \Psi^{(z)}\9_{2} +\nonumber 
|\Phi^-\9_{01} |\tilde \Psi^{(x)}\9_{2}+ |\Phi^+\9_{01} |\tilde
\Psi^{(y)}\9_{2}\right), \label{ident}
\eeq
\end{widetext}
where the four Bell states are given by Eqs.~(\ref{psipm}) and
(\ref{phipm}), and the symbol $|\tilde \Psi^{(z)}\9$ means the
state $|\Psi\9$ rotated by $\pi$ around the $z$-axis, etc. Thus,
the Bell measurement performed on the two particles at Alice's
site leads to one of the branches of the superposition on the rhs
of Eq.~(\ref{ident}). To complete the teleportation, Bob performs
a rotation  by $\pi$ around one of the axes according to the
classical information he gets from Alice.

Gauge theories also lead to interesting questions about
measurability. Wilson loops, which are nonlocal objects by
definition, are often invoked in their presentation (Peskin and
Schroeder, 1995) and are the backbone of lattice gauge theories
(Makeenko, 2002). Beckman \etal\ (2002) investigated the
measurability of the Wilson loop operators.

The impossibility of instantaneous communication allows to
circumvent the theoretical impossibility of quantum bit commitment
(Mayers, 1997; Lo and Chau, 1997). Kent (1999, 2003) developed
protocols based on the finite speed of communication and evaluated
their communication costs and security. In particular Kent's RBC2
protocol allows a bit commitment to be indefinitely maintained
with unconditional security against all classical attacks, and at
least for some finite amount of time against quantum attacks
(Kent, 2003).

\section{The relativistic measuring process}

\subsection{General properties}

Quantum measurements are usually considered as quasi-instantaneous
processes. In particular, they affect the wave function
instantaneously throughout the entire configuration space.
Measurements of finite duration (Peres and Wootters, 1985) make no
essential difference in this respect.  Is this quasi-instantaneous
change of the quantum state, caused by a local intervention of an
exophysical agent, consistent with relativity theory? The answer
is not obvious. The wave function itself is not a material object
forbidden to travel faster than light, but we may still ask how
the dynamical evolution of an extended quantum system that
undergoes several measurements in distant spacetime regions is
described in different Lorentz frames.

Difficulties were pointed out long ago by Bloch (1967), Aharonov
and Albert (1981, 1984), and many others (Peres, 1995 and
references therein). Still before them, in the very early years of
quantum mechanics, Bohr and Rosenfeld (1933) had given a complete
relativistic theory of the measurement of quantum {\it fields\/},
but these authors were not concerned about the properties of the
new quantum states that resulted from these measurements and their
work does not answer the question that was raised above. Other
authors (Scarani \etal, 2000; Zbinden \etal, 2001) considered
detectors in relative motion, and therefore at rest in different
Lorentz frames.  These works also do not give an explicit answer
to the above question: a detector in uniform motion is just as
good as one that has undergone an ordinary spatial rotation.
(Accelerated detectors involve new physical phenomena, see
Sec.~\ref{unruh}.)  The point is not how individual detectors
happen to move, but how the effects due to these detectors are
described in different ways in one Lorentz frame or another.

To become fully relativistic, the notion of intervention requires
some refinement. The precise location of an intervention, which is
important in a relativistic discussion, is the point from which
classical information is sent that may affect the input of other
interventions. More precisely, it is the earliest small region of
spacetime from which classical information could have been sent.
Moreover, in the conventional presentation of non-relativistic
quantum mechanics, each intervention has a (finite) number of
outcomes, for example, this or that detector clicks. In a
relativistic treatment, the spatial separation of the detectors is
essential and each detector corresponds to a different
intervention. The reason is that if several detectors are set up
so that they act at a given time in one Lorentz frame, they would
act at different times in another Lorentz frame. However, a
knowledge of the time ordering of events is essential in our
dynamical calculations, so that we want the parameters of an
intervention to refer unambiguously to only one time (indeed to
only one spacetime ``point''). Therefore, an intervention can
involve only one detector and it can have only two possible
outcomes: either there was a ``click'' or there wasn't.

What is the role of relativity theory here? We may likewise ask
what is the role of translation and/or rotation invariance in a
nonrelativistic theory. The point is that the rules for computing
quantum probabilities involve explicitly the spacetime coordinates
of the interventions. Lorentz invariance (or rotational
invariance, as a special case) says that if the classical
spacetime coordinates are subjected to a particular linear
transformation, then the probabilities remain the same. This
invariance is not trivial because the rule for computing the
probability of occurrence of a given record involves a sequence of
mathematical operations corresponding to the time ordered set of
all the relevant interventions.

If we only consider the Euclidean group, all we have to know is
how to transform the classical parameters, and the wave function,
and the various operators, under translations and rotations of the
coordinates. However, when we consider genuine Lorentz
transformations, we have not only to Lorentz-transform the above
symbols, but we are faced with a new problem: the natural way of
calculating  the result of a sequence of interventions, namely by
considering them in chronological order, is different for
different inertial frames. The issue is not only a matter of
covariance of the symbols at each intervention and between
consecutive interventions.  There are genuinely different
prescriptions for choosing the sequence of mathematical operations
in our calculation. Therefore these different orderings ought to
give the same set of probabilities, and this demand is not
trivial.

$ $

\begin{figure}[htbp]
\epsfxsize=0.44\textwidth
\centerline{\epsffile{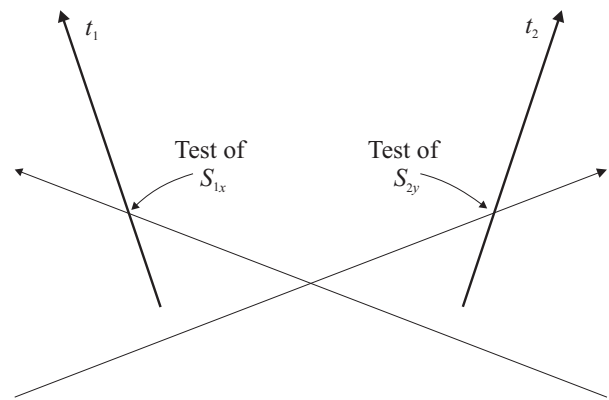}} \vspace*{0.1cm} \caption{\small{In this
spacetime diagram, the origins of the
coordinate systems are the locations of the two tests. The $t_1$
and $t_2$ axes are the world lines of the observers, who are
receding from each other. In each Lorentz frame, the $z_1$ and
$z_2$ axes are isochronous: \,$t_1=0$ \,and \,$t_2=0$,
respectively.}}
\end{figure}

\subsection{The role of relativity}

A typical example of relativistic measurement is the detection
system in the experimental facility of a modern high energy
accelerator. Following a high energy collision, thousands of
detection events occur in locations that may be mutually
space-like.  Yet, some of the detection events are mutually
time-like, for example when the world line of a charged particle
is recorded in an array of wire chambers. In a relativistic
context, the term ``detector'' strictly means an elementary
detecting element, such as a bubble in a bubble chamber, or a
small segment of wire in a wire chamber.\footnote{High energy
physicists use a different language. For them, an ``event'' is one
high energy collision together with all the subsequent detections
that are recorded. This ``event'' is what we call here an
experiment (while they call ``experiment'' the complete
experimental setup that may be run for many months). And their
``detector'' is a huge machine weighing thousands of tons.}

A much simpler example of space-like separated interventions,
which is amenable to a complete analysis, is Bohm's version of the
Einstein-Podolsky-Rosen ``paradox''  (hereafter EPRB; Einstein,
Podolsky, and Rosen, 1935; Bohm 1951) which is sketched in Fig.~1,
with two coordinate systems in relative motion (Peres, 1993). In
that experiment, a pair of spin-$1\over2$ particles, prepared in a
singlet state, move apart and are detected by two observers. Each
observer measures a spin component along an arbitrarily chosen
direction. The two interventions are mutually space-like as shown
in the figure. The test of $S_{1x}$ occurs first when recorded in
$t_1$-time, and the test of $S_{2y}$ is the first one in
$t_2$-time. The evolution of the quantum state of this bipartite
system appears to be genuinely different when recorded in two
Lorentz frames in relative motion. The quantum states are not
Lorentz-transforms of each other. Yet, all the observable results
are the same. Consistency of the theoretical formalism imposes
definite relationships between the various operators used in the
calculations (Peres, 2000b). In particular, it is sufficient for
consistency that the Kraus operators satisfy {\it equal-time\/}
commutation relation as in Eq.~(\ref{etcr}). The analogy with
relativistic quantum field theory is manifest.

In general, consider the quantum evolution from an initial state
$\rho_0$ to a final state $\rho_f$. It is a completely positive
map,
\beq
\rho_f=\sum_n A_n\,\rho_0\,A^\dagger_n.
\eeq
The Lorentz transformation of the Kraus matrices $A_n$ can be
obtained as follows. We have $\rho_0'=U\rho_0 U^\dagger$ and
$\rho_f'=V\rho_f V^\dagger$, where $U$ and $V$ are unitary
representations of Lorentz transformations for the systems
represented by $\rho_0$ and $\rho_f$ (which may be of different
nature and even of different dimensions).

Lorentz invariance means that, in another frame, the Kraus
matrices $A_n'$ satisfy
\beq
\rho'_f=\sum_n A'_n\,\rho'_0\,A'^\dagger_n.
\eeq
A simple solution is
\beq
A'_n=VA_n\,U^\dagger,
\eeq
but this is not the most general one. The latter is
\beq
A'_n=\sum_m W_n^m\,VA_m\,U^\dagger,
\eeq
where $W_n^m$ is a unitary matrix that acts on the labels $m,n$
(not on the Hilbert spaces of $\rho_0$ and $\rho_f$). This
arbitrariness is a kind of gauge freedom, and can be resolved only
by a complete dynamical description of the intervention process.
This, however, is an arduous problem. Relativistic interactions
necessarily involve field theory, and the question is how to
generalize the quantum information tools (POVMs, completely
positive maps) into objects that are described by quantum field
theories (Terno 2002).

At this stage we consider only field theories in Minkowski
spacetime where a unique vacuum state $|\Omega\9$ is defined. The
discrete indices that appear in the above equations can still be
used, owing to the fact that the underlying Hilbert space is
separable (Streater and Wightman, 1964). Therefore the formalism
is valid without change in  the relativistic domain.\footnote{The
fact that the values of classical parameters (``measurable
quantities'') are finite real numbers is sufficient to construct
probability measures. For the exact formulation see Davies (1976)
and Holevo (1982). Similar arguments justify the inclusion of only
bounded operators into algebras of local observables (Haag, 1996;
Araki, 1999).} However, not every measurement-induced state
transformation that can be written in the Kraus form is permitted
or makes sense. Relativity theory prohibits superluminal velocity
for material objects. Consistency with the requirements of
covariance and causality is an intrinsic feature of quantum field
theories. Nevertheless, to make problems solvable, a patchwork of
relativistic and non-relativistic theories is employed. For
example, a measurement on relativistic systems is usually treated
by introducing detectors that are described by non-relativistic
quantum mechanics. Often these detectors are stripped to only a
few discrete degrees of freedom (Unruh and Wald, 1984; Levin,
Peleg and Peres, 1992; Wald, 1994).

An external probe which is not described by field theory and whose
coupling to the fields of interest is arbitrarily adjustable is
obviously an idealization. Beckman \etal\ (2001) assert that if
the probe variables are ``heavy,'' with rapidly decaying
correlations and the field variables are ``light,'' then this
idealization is credible. Still, causality requirements like the
absence of signalling should be checked for any proposed
measurement scheme (Sec.~\ref{no-com} also discusses causality
requirements).

Consider again the descriptions of the EPRB gedankenexperiment in
two coordinate systems in relative motion. There exists a Lorentz
transformation connecting the initial states $\rho_0$ and
$\rho_0'$ before the two interventions, and likewise there is a
Lorentz transformation connecting the final states $\rho_f$ and
$\rho_f'$ after completion of the two interventions. On the other
hand, there is no Lorentz transformation relating the states at
intermediate times that are not in the past or future of {\it
both\/} interventions (Peres, 2000b). The various Kraus operators,
acting at different times, appear in different orders.
Nevertheless the overall transition from initial to final state is
Lorentz invariant (Peres, 2001).

In the time interval between the two interventions, {\it nothing}
actually happens in the real world. It is only in our mathematical
calculations that there is a deterministic evolution of the state
of the quantum system. This evolution is {\it not\/} a physical
process.\footnote{Likewise, the quantum state of Schr\"odinger's
legendary cat, doomed to be killed by an automatic device
triggered by the decay of a radio\-active atom, evolves into a
superposition of ``live'' and ``dead'' states. This is a
manifestly absurd situation for a real cat. The only meaning that
such a quantum state can have is that of a mathematical tool for
statistical predictions on the fates of numerous cats subjected to
the same cruel experiment.} What distinguishes the intermediate
evolution {\it between\/} interventions from the one occurring
{\it at\/} an intervention is the unpredictability of the outcome
of the latter: either there is a click or there is no click of the
detector. This unpredictable macroscopic event starts a new
chapter in the history of the quantum system which acquires a new
state, according to Eq.~(\ref{ArhoA}).

\subsection{Quantum nonlocality?}

Phenomena like those illustrated in Fig.~1 are often attributed to
``quantum nonlocality'' and have led some authors to speculate on
the possibility of superluminal communication (actually,
instantaneous communication). One of these proposals (Herbert,
1981) looked reasonably serious and arose enough interest to lead
to investigations disproving this possibility (Glauber, 1986) and
in particular to the discovery of the no-cloning theorem (Wootters
and Zurek, 1982; Dieks, 1982). Let us examine more closely the
origin of these claims of nonlocality.

Bell's theorem (1964) asserts that it is impossible to mimic
quantum theory by introducing a set of objective {\it local\/}
``hidden'' variables.  It follows that any classical imitation of
quantum mechanics is necessarily nonlocal. However Bell's theorem
does not imply the existence of any nonlocality in quantum theory
itself. In particular relativistic quantum field theory is
manifestly local.  The simple and obvious fact is that information
has to be carried by {\it material objects\/}, quantized or not.
Therefore quantum measurements do not allow any information to be
transmitted faster than the characteristic velocity that appears
in the Green's functions of the particles emitted in the
experiment. In a Lorentz invariant theory, this limit is the
velocity of light.

In summary, relativistic causality cannot be violated by quantum
measurements. The only physical assumption that is needed to prove
this assertion is that Lorentz transformations of the spacetime
coordinates are implemented in quantum theory by {\it unitary\/}
transformations of the various operators. This is the same as
saying that the Lorentz group is a valid symmetry of the physical
system (Weinberg, 1995).

\subsection{Classical analogies}

Are relativity and quantum theory really involved in these issues?
The matter of information transfer by means of distant
measurements is essentially nonrelativistic.  Replace
``superluminal'' by ``supersonic'' and the argument is exactly the
same. The maximal speed of communication is determined by the
dynamical laws that govern the physical infrastructure. In quantum
field theory, the field excitations are called ``particles'' and
their speed over macroscopic distances cannot exceed the speed of
light. In condensed matter physics, linear excitations are called
phonons and the maximal speed is that of sound.

As to the EPRB setup, consider an analogous  classical situation:
a bomb, initially at rest, explodes into two fragments carrying
opposite angular momenta. Alice and Bob, far away from each other,
measure arbitrarily chosen components of ${\bf J}_1$ and ${\bf
J}_2$. (They can measure all the components, since these have
objective values.) Yet, Bob's measurement tells him nothing of
what Alice did, nor even whether she did anything at all. He can
only know with certainty what {\it would\/} be the result found by
Alice {\it if\/} she measures her {\bf J} along the same direction
as him, and make statistical inferences for other possible
directions of Alice's measurement.

The classical-quantum analogy becomes complete if we use classical
statistical mechanics. The distribution of bomb fragments is given
by a Liouville function in phase space. When Alice measures ${\bf
J}_1$, the Liouville function for ${\bf J}_2$ is instantly
altered, however far Bob is from Alice. No one finds this
surprising, since it is universally agreed that a Liouville
function is only a mathematical tool representing our statistical
knowledge.  Likewise, the wave function $\psi$, or the
corresponding Wigner function (Wigner, 1932) which is the quantum
analogue of a Liouville function, are no more than mathematical
tools for computing probabilities. It is only when they are
regarded as physical objects that superluminal paradoxes arise.

The essential difference between the classical and quantum
functions which change instantaneously as the result of
measurements is that the classical Liouville function is attached
to objective properties that are only imperfectly known. On the
other hand, in the quantum case, the probabilities are attached to
{\it potential\/} outcomes of mutually incompatible experiments,
and these outcomes do not exist ``out there'' without the actual
interventions. Unperformed experiments have no results.

\section{Quantum entropy and special relativity}

\subsection{Reduced density matrices}

In our discussion of the measuring process, decoherence was
attributed to the unability of accounting explicitly for the
degrees of freedom of the environment. The environment thus
behaves an exosystem (Finkelstein, 1988) and the system of
interest is ``open'' because parts of the universe are excluded
from its description.

This leads to the introduction of reduced density matrices: let us
use Latin indices for the description of the exosystem (that is,
if we were able to give it a description) and Greek indices for
the subsystem that we can actually describe. The components of a
state vector would thus be written $V_{m\mu}$ and those of a
density matrix $\rho_{m\mu,n\nu}$. The reduced density matrix of
the system of interest is given by
\beq
\tau_{\mu\nu}=\sum_{m} \rho_{m\mu,m\nu}. \label{tau}
\eeq
Even if $\rho$ is a pure state (a matrix of rank one), $\tau$ is
in general a mixed state. Its {\it entropy\/} is defined as
\beq
S=-\tr(\tau\log\tau).
\eeq

In a relativistic system, whatever is outside the past light cone
of the observer is unknown to him, but also cannot affect his
system, therefore does not lead to decoherence (here, we assume
that no particle emitted by an exosystem located outside the past
cone penetrates into the future cone.) Since observers located at
different points have different past light cones, they exclude
from their descriptions different parts of spacetime. Therefore
any transformation law between them must tacitly assume that the
part excluded by one observer is irrelevant to the system of the
other observer.

Another consequence of relativity is that there is a hierarchy of
dynamical variables: {\it primary variables\/} have relativistic
transformation laws that depend only on the Lorentz transformation
matrix $\Lambda$ that acts on the spacetime coordinates. For
example, momentum components are primary variables. On the other
hand, {\it secondary variables\/} such as spin and polarization
have transformation laws that depend not only on $\Lambda$, but
also on the momentum of the particle. As a consequence, the
reduced density matrix for secondary variables, which may be well
defined in any coordinate system, has no transformation law
relating its values in different Lorentz frames. A simple example
is given in Sec.~\ref{massive}. Appendix A gives a summary of the
relativistic state transformations for free particles.

Moreover, an unambiguous definition of the reduced density matrix
by means of Eq.~(\ref{tau}) is possible only if the secondary
variables are unconstrained. For gauge field theories, that
equation may be meaningless if it conflicts with constraints
imposed on the physical states (Beckman \etal, 2002; Peres and
Terno, 2003). In the absence of a general prescription, a
case-by-case treatment is required. A particular construction,
valid with respect to a certain class of tests, is given in
Sec.~\ref{photons}.  A general way of defining reduced density
matrices for physical states in gauge theories is an open problem.

\subsection{Massive particles} \label{massive}

We first consider the relativistic properties of the spin entropy
for a single, free particle of spin~\half\ and mass $m>0$. We
shall show that the usual definition of quantum entropy has no
invariant meaning. The reason is that under a Lorentz boost, the
spin undergoes a Wigner rotation (Wigner, 1939; Halpern, 1968)
whose direction and magnitude depend on the momentum of the
particle. Even if the initial state is a direct product of a
function of momentum and a function of spin, the transformed state
is not a direct product. Spin and momentum appear to be entangled.
(This is not the familiar type of entanglement which can be used
for quantum communication, because both degrees of freedom belong
to the same particle, not to distinct subsystems that could be
widely separated.)

The quantum state of a spin-\half\ particle can be written, in the
momentum representation, as a two-component spinor,
\beq
\psi(\pp)={a_1(\pp)\choose a_2(\pp)}, \label{psi0}
\eeq
where the amplitudes $a_r$ satisfy $\sum_r\int|a_r(\pp)|^2
d\pp=1$. The normalization of these amplitudes is a matter of
convenience, depending on whether we prefer to include a factor
\mbox{$p_0=(m^2+\pp^2)^{1/2}$} in it, or to have such factors in
the transformation law (\ref{tf}) below.  Following Halpern
(1968), we shall use the second alternative, because it is closer
to the nonrelativistic notation which appears in the usual
definition of entropy. In this section, we use natural units:
$c=1$.

Here we emphasize that we consider normalizable states, in the
momentum representation, not momentum eigenstates as usual in
textbooks on particle physics. The latter are chiefly concerned
with the computation of $\6\mbox{in}|\mbox{out}\9$ matrix elements
needed to obtain cross sections and other asymptotic properties.
However, in general a particle has no definite momentum. For
example, if an electron is elastically scattered by some target,
the electron state after the scattering is a superposition that
involves momenta in all directions.

In that case, it still is formally possible to ask, in any Lorentz
frame, what is the value of a spin component in a given direction
(this is a legitimate Hermitian operator). In quantum information
theory, the important issue does not reside in asymptotic
properties, but how entanglement (a communication resource) is
defined by different observers. Early papers on this subject used
momentum eigenstates, just as in particle physics (Czachor, 1997).
However, radically new properties arise when localized quantum
states are considered.

Let us define a reduced density matrix, $\tau=\int
d\pp\,\psi(\pp)\psi^\dagger(\pp)$, giving statistical predictions
for the results of measurements of spin components by an ideal
apparatus which is not affected by the momentum of the particle.
The spin entropy is
\beq
S=-\tr(\tau\log\tau)=-\sum\lambda_j\log\lambda_j,
\eeq
where $\lambda_j$ are the eigenvalues of $\tau$.

As usual, ignoring some degrees of freedom leaves the others in a
mixed state. What is not obvious is that in the present case the
amount of mixing depends on the Lorentz frame used by the
observer.  Indeed consider another observer (Bob) who moves with a
constant velocity with respect to Alice who prepared state
(\ref{psi0}). In the Lorentz frame where Bob is at rest, the same
spin-\half\ particle has a state
\beq
\psi'(\pp)={a'_1(\pp)\choose a'_2(\pp)}.
\eeq
The transformation law is (Weinberg, 1995)
\beq
a'(\pp)=[(\Lambda^{-1}p)_0/p_0]^{1/2}\,\sum_s
 D_{rs}[\Lambda,(\Lambda^{-1}p)]\,a_s(\Lambda^{-1}p) \label{tf},
 \eeq
where $D_{rs}$ is the Wigner rotation matrix for a Lorentz
transformation $\Lambda$. Further details of this transformation
and its representation by a quantum circuit are given in Appendix
A.

As an example, take a particle prepared by Alice with spin in the
$z$ direction, so that $a_2(\pp)=0$. Spin and momentum are not
entangled, and the spin entropy is zero. When that particle is
described in Bob's Lorentz frame, moving with velocity $\beta$ in
a direction at an angle $\theta$ with Alice's $z$-axis, a detailed
calculation shows that both $a'_1$ and $a'_2$ are nonzero, so that
the spin entropy is positive (Peres, Scudo, and Terno, 2002). This
phenomenon is illustrated in Fig.~2. A relevant parameter, apart
from the angle $\theta$, is, in the leading order in momentum
spread,
\beq
\Gamma=\frac{\Delta}{m}\,\frac{1-\sqrt{1-\beta^2}}{\beta},
\eeq
where $\Delta$ is the momentum spread in Alice's frame. {\it The
entropy has no invariant meaning\/}, because the reduced density
matrix $\tau$ has no covariant transformation law, except in the
limiting case of sharp momenta.  Only the complete density matrix
transforms covariantly.

How is the linearity of the transformation laws lost in this
purely quantum mechanical problem? The momenta \pp\ do transform
linearly, but the law of transformation of spin depends explicitly
on~\pp. When we evaluate $\tau$ by summing over momenta in $\rho$,
all knowledge of these momenta is lost and it is then impossible
to obtain $\tau'$ by transforming $\tau$. Not only is linearity
lost, but the result is not nonlinearity in the usual sense of
this term. It is the absence of {\it any\/} definite
transformation law which depends only on the Lorentz matrix.

It is noteworthy that a similar situation arises for a classical
system whose state is given in any Lorentz frame by a Liouville
function (Balescu and Kotera, 1967). Recall that a Liouville
function expresses our probabilistic description of a classical
system --- what we can predict before we perform an actual
observation --- just as a quantum state is a mathematical
expression used for computing probabilities of events.

To avoid any misunderstanding, we emphasize that there is no
consistent relativistic statistical mechanics for $N$ interacting
particles, with a $6N$-dimensional phase space defined by the
canonical coordinates ${\bf p}_n$ and ${\bf q}_n$
($n=1,\ldots,N$). Any relativistic interaction must be mediated by
{\it fields\/}, having an infinity of degrees of freedom. A
complete Liouville function, or rather Liouville functional, must
therefore contain not only all the canonical variables ${\bf p}_n$
and ${\bf q}_n$, but also all the fields. However, once this
Liouville functional is known (in principle), we can define from
it a reduced Liouville function, by integrating the functional
over all the degrees of freedom of the fields. The result is a
function of ${\bf p}_n$ and ${\bf q}_n$ only (just as we compute
reduced density matrices in quantum theory). The time evolution of
such reduced Liouville functions cannot be obtained directly from
canonical Hamiltonian dynamics without explicitly mentioning the
fields. These functions are well defined in any Lorentz frame, but
they have no relativistic transformation law. Only the complete
Liouville functional, including the fields, has one.

Consider now a pair of orthogonal states that were prepared by
Alice. How well can moving Bob distinguish them, if he is
restricted to measuring discrete degrees of freedom? We shall use
the simplest criterion, namely the probability of error $P_E$,
defined as follows: an observer receives a single copy of one of
the two known states and performs any operation permitted by
quantum theory in order to decide which state was supplied. The
probability of a wrong answer for an optimal measurement is (Fuchs
and van de Graaf, 1999)
\beq
P_E(\rho_1,\rho_2)=\half+\mbox{$1\over4$}\,{\rm tr}
 \sqrt{(\rho_1-\rho_2)^2}.  \label{pe}
 \eeq
In Alice's frame $P_E=0$. It can be shown that in Bob's frame,
$P'_E\propto\Gamma^2$, where the proportionality factor depends on
the angle $\theta$ defined above. Of course, the opposite Lorentz
transformation induces a  change from a positive $P_E$ in Bob's
frame to $P_E$=0 in Alice's frame. We discuss the resulting
effective quantum channel in Sec.~\ref{cocha}.

\begin{figure}[htbp]
\epsfxsize=0.46\textwidth
\centerline{\epsffile{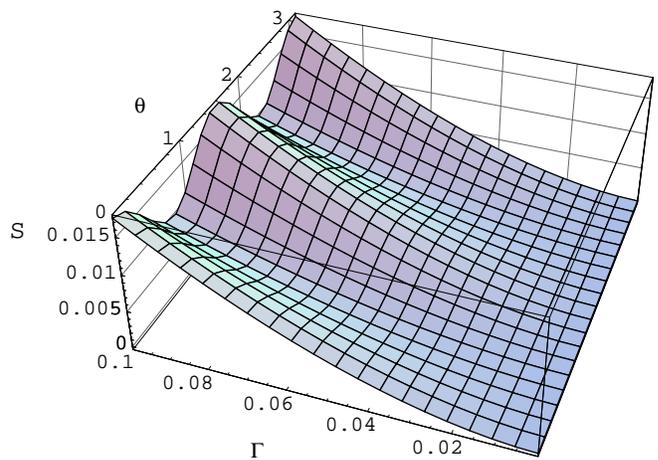}} \vspace*{-0.1cm} \caption{\small{
Dependence of the spin entropy $S$, in
Bob's frame, on the values of the angle $\theta$ and a parameter
$\Gamma\approx[1-(1-\beta^2)^{1/2}]\Delta/m\beta$, where $\Delta$
is the momentum spread in Alice's frame.}}
\end{figure}

\subsection{Photons}\label{photons}
The long range propagation of polarized photons is an essential
tool of quantum crypto\-graphy (Gisin \etal, 2002). Usually,
optical fibers are used, and the photons may be absorbed or
depolarized due to imperfections. In some cases, such as
communication with space stations, the photons  propagate in vacuo
(Buttler \etal, 2000).  The beam then has a finite diffraction
angle of order $\lambda/a$, where $a$ is the aperture size, and
new deleterious effects appear. In particular a polarization
detector cannot be rigorously perpendicular to the wave vector and
the transmission is never faithful, even with perfect detectors.
Moreover, this ``vacuum noise'' depends on the relative motion of
the observer with respect to the source.

These relativistic effects are essentially different from those
for massive particles that were discussed above, because photons
have only two linearly independent polarization states. The
properties that we discuss are kinematical, not dynamical. At the
statistical level, it is not even necessary to involve quantum
electrodynamics. Most formulas can be derived by elementary
classical methods (Peres and Terno, 2003). It is only when we
consider individual photons, for crypto\-graphic applications,
that quantum theory becomes essential. The diffraction effects
mentioned above lead to superselection rules which make it
impossible to define a reduced density matrix for polarization. As
shown below, it is still possible to have ``effective'' density
matrices; however, the latter depend not only on the preparation
process, but also on the method of detection that is used by the
observer.

Assume for simplicity that the electromagnetic signal is
monochromatic. In a Fourier decomposition, the Cartesian
components of the wave vector $k_\mu$ (with $\mu=0,1,2,3$) can be
written in term of polar angles:
\beq
k_\mu=(1,\sin\theta\cos\phi,\sin\theta\sin\phi,\cos\theta),
\eeq
where we use units such that $c=1$ and $k_0=1$. Let us choose the
$z$ axis so that a well collimated beam has a large amplitude only
for small $\theta$.

In a real experiment, the angles $\theta$ and $\phi$ are
distributed in a continuous way around the $z$ axis (exactly how
depends on the properties of the laser) and one has to take a
suitable average over them. As the definition of polarization
explicitly depends on the direction of \bk,  taking the average
over many values of \bk\ leads to an impure polarization and may
cause transmission errors.

Let us consider the effect of a motion of the detector relative to
the emitter, with a constant velocity ${\bf v}=(0,0,v)$. The
Lorentz transformation of $k_\mu$ in Eq.~(\theequation) yields new
components
\beq
k'_0=\gamma(1-v\cos\theta)\qquad\mbox{and}\qquad
  k'_z=\gamma(\cos\theta-v),  \label{gamma}
\eeq
where $\gamma=(1-v^2)^{-1/2}$. Considering again a single Fourier
component, we have, instead of the unit vector \bk, a new unit
vector
\beq
\bk'=\left(\frac{\sin\theta}{\gamma(1-v\cos\theta)},\;0,\;
  \frac{\cos\theta-v}{1-v\cos\theta}\right).\label{loren1}
  \eeq
In other words, there is a new tilt angle $\theta'$ given by
\beq
\sin\theta'=\sin\theta/\gamma(1-v\cos\theta).
\eeq
For small $\theta$, such that $\theta^2\ll|v|$, we have
\beq
\theta'=\theta \sqrt{{1+v}\over{1-v}}. \label{doppler}
\eeq
The square root is the familiar relativistic Doppler factor. For
large negative $v$, the diffraction angle becomes arbitrarily
small, and sideway losses (which are proportional to $\theta'^2$)
can be reduced to zero.

It is noteworthy that the same Doppler factor was obtained by
Jarett and Cover (1981) who considered only the relativistic
transformations of bit rate and noise intensity, without any
specific physical model. This remarkable agreement shows that
information theory should properly be considered as a branch of
physics.

In applications to secure communication, the ideal scenario is
that isolated photons (single particle Fock states) are emitted.
In a more realistic setup, the transmission is by means of weak
coherent pulses containing on the average less than one photon
each.  A basis of the one-photon space is spanned by states of
definite momentum and helicity,
\beq
|\bk,\bep_\bk^\pm\9 \equiv |\bk\9\otimes|\bep_\bk^\pm\9,
 \label{basis}
 \eeq
where the momentum basis is normalized by $\6\bq|\bk\9=(2\pi)^3(2
k^0)\delta^{(3)}(\bq-\bk)$, and helicity states $|\bep_\bk^\pm\9$
are explicitly defined by Eq.~(\ref{helivectors}) below.

As we know, polarization is a {\it secondary variable\/}: states
that correspond to different momenta belong to distinct Hilbert
spaces and cannot be superposed (an expression such as
$|\bep_\bk^\pm\9+|\bep_\bq^\pm\9$ is meaningless if $\bk\neq\bq$).
The complete basis (\ref{basis}) does not violate this
superselection rule, owing to the othogonality of the momentum
basis. Therefore, a generic one-photon state is given by a wave
packet
\beq
|\Psi\9=\int d\mu(\bk)f(\bk)|\bk,\bal(\bk)\9.\label{photon}
\eeq
The Lorentz-invariant measure is $d\mu(k)=d^3\bk/(2\pi)^3 2k^0$,
and normalized states satisfy $\int d\mu(k)|f(\bk)|^2=1$.  The
generic polarization state $|\bal(\bk)\9$ corresponds to the
geometrical 3-vector
\beq
\bal(\bk)=\alpha_+(\bk)\bep^+_\bk+\alpha_-(\bk)\bep^-_\bk,
 \label{elliptic}
 \eeq
where $|\alpha_+|^2+|\alpha_-|^2=1$, and the explicit form of
$\bep^\pm_\bk$ is given below.

Lorentz transformations of quantum states are most easily computed
by referring to some standard momentum, which for photons is
$p^\nu=(1,0,0,1)$. Accordingly, standard right and left circular
polarization vectors are $\bep^\pm_p=(1,\pm i,0)/\sqrt{2}$. For
{\it linear\/} polarization, we take Eq.~(\ref{elliptic}) with
$\alpha_+=(\alpha_-)^*$, so that the 3-vectors $\bal(\bk)$ are
real. In general, complex $\bal(\bk)$ correspond to elliptic
polarization.

Under a Lorentz transformation $\Lambda$, these states become
$|\bk_\Lambda,\bal(\bk_\Lambda)\9$, where $\bk_\Lambda$ is the
spatial part of a four-vector $k_\Lambda=\Lambda k$, and the new
polarization vector can be obtained by an appropriate rotation
given by Eq.~(\ref{heltra}) below.  For each $\bk$ a polarization
basis consists of the helicity vectors,
\beq
\bep^\pm_\bk=R(\hbk)\bep^\pm_p, \label{helivectors}
\eeq
and the corresponding quantum states are $|\bk,\bep^\pm_\bk\9$. As
usual, $\hbk$ denotes the unit 3-vector in the direction of $\bk$.
The {\it standard} matrix (Weinberg, 1995) that rotates the
standard direction $(0,0,1)$ to
$\hbk=(\sin\theta\cos\phi,\sin\theta\sin\phi,\cos\theta)$  is
\beq
R(\hbk)=\left(\bay{ccccc}
\cos\theta\cos\phi & & -\sin\phi  & & \cos\phi\sin\theta \\
\cos\theta\sin\phi & & \cos\phi & & \sin\phi\sin\theta \\
-\sin\theta & & 0 & & \cos\theta
\eay\right),
\eeq
and likewise for $\hbk_\Lambda$.

Under a general Lorentz transformation, be it a rotation or a
boost, helicity is preserved, but quantum states and the
corresponding geometric vectors acquire helicity-dependent phases
(see Appendix A for more details):
\beq
\alpha_+\bep^+_\bk+\alpha_-\bep^-_\bk\rightarrow \alpha_+
e^{i\xi(\Lambda,\hbk)}\bep^+_{\bk_\Lambda}+
\alpha_-e^{-i\xi(\Lambda,\hbk)}\bep^-_{\bk_\Lambda},
\label{heltra}
\eeq
where the explicit expressions for $\xi(\Lambda,\hbk)$ are given
by Lindner, Peres, and Terno (2003) and Bergou, Gingrich and Adami
(2003).

 The superselection rule that was mentioned above makes it
impossible to define a reduced density matrix in the usual way
(Peres and Terno, 2003; Lindner, Peres and Terno, 2003). We can
however define an ``effective'' reduced density matrix for
polarization, as follows. The labelling of polarization states by
Euclidean vectors $\be_\bk^n$, and the fact that photons are
spin-1 particles, suggest the use of a $3\times 3$ matrix with
entries labelled $x$, $y$ and $z$. Classically, they correspond to
different directions of the electric field. For example, when
$\hbk=\hat{\bf z}$, only $\rho_{xx}$, $\rho_{xy}$, $\rho_{yy}$ are
non-zero. For a generic photon state $|\Psi\9$, let us try to
construct a reduced density matrix $\rho_{xx}$ that gives the
expectation value of an operator representing the polarization in
the $x$ direction, irrespective of the particle's momentum.

To have a momentum-independent polarization is to tacitly admit
longitudinal photons. Unphysical concepts are often used in
intermediate steps in theoretical physics. Momentum-independent
polarization states thus consist of physical (transversal) and
unphysical (longitudinal) parts, the latter corresponding to a
polarization vector $\bep^\ell=\hbk$. For example, a generalized
polarization state along the $x$-axis is
\beq
|\hbx\9=x_+(\bk)|\bep^+_\bk\9+x_-(\bk)|\bep^-_\bk\9+
x_\ell(\bk)|\bep^\ell_\bk\9,\label{decomp}
\eeq
where $x_\pm(\bk)=\bep^\pm_\bk\cdot\hbx$, and $x_\ell(\bk)=
\hbx\cdot\hbk=\sin\theta\cos\phi$. It follows that
$|x_+|^2+|x_-|^2+|x_\ell|^2=1$, and we thus define
\beq
\be_x(\bk)=\frac{x_+(\bk)\bep^+_\bk+x_-(\bk)\bep^-_\bk}
{\sqrt{x_+^2+x_-^2}} \label{physdir},
\eeq
as the polarization vector associated with the $x$ direction. It
follows from (\ref{decomp}) that $\6\hbx|\hbx\9=1$ and
$\6\hbx|\hby\9=\hbx\cdot\hby=0$, and likewise for other
directions, so that
\beq
|\hbx\9\6\hbx|+|\hby\9\6\hby|+|\hbz\9\6\hbz|=\1\label{xyz}.
\eeq

To the direction $\hbx$ corresponds a projection operator
\beq
P_{x}=|\hbx\9\6\hbx|\otimes \1_p=|\hbx\9\6\hbx|\otimes \int
d\mu(\bk)|\bk\9\6\bk|,
\eeq
where $\1_p$ is the unit operator in momentum space. The action of
$P_{x}$ on $|\Psi\9$ follows from Eq.~(\ref{decomp}) and
$\6\bep^\pm_\bk|\bep^\ell_\bk\9=0$. Only the transversal part of
$|\hbx\9$ appears in the expectation value:
\beq
\6\Psi|P_{x}|\Psi\9=\int d\mu(\bk)|f(\bk)|^2|x_
 +(\bk)\alpha_+^*(\bk)+x_-(\bk)\alpha_-^*(\bk)|^2.
 \eeq
It is convenient to write the transversal part of $|\hbx\9$ as
\begin{eqnarray}
|\bb_x(\bk)\9 & \equiv &
 (|\bep^+_\bk\9\6\bep^+_\bk|+|\bep^-_\bk\9\6\bep^-_\bk|)|\hbx\9,\\
 &=&  x_+(\bk)|\bep^+_\bk\9+x_-(\bk)|\bep^-_\bk\9.
\label{vector}
\end{eqnarray}
Likewise define  $|\bb_y(\bk)\9$ and $|\bb_z(\bk)\9$. These three
state vectors are neither of unit length nor mutually orthogonal.
For $\bk= (\sin\theta\cos\phi,\sin\theta\sin\phi,\cos\theta)$ we
have
\begin{eqnarray}
|\bb_x(\bk)\9 =
\s12(\cos\theta\cos\phi+i\sin\phi)|\bep^+_\bk\9+ \\
 \s12(\cos\theta\cos\phi-i\sin\phi)|\bep^-_\bk\9
   \equiv   c(\theta,\phi)|\bk,\be_x(\bk)\9,
\end{eqnarray}
where $\be_x(\bk)$ is given by Eq.~(\ref{physdir}),  and
$c(\theta,\phi)=\sqrt{x_+^2+x_-^2}$.

Finally, a POVM element $E_{x}$ which is the physical part of
$P_{x}$, namely is equivalent to $P_{x}$ for physical states
(without longitudinal photons) is
\beq
E_{x}=\int
d\mu(\bk)|\bk,\bb_x(\bk)\9\6\bk,\bb_x(\bk)|,\label{povmexa}
\eeq
and likewise for the other directions. The operators $E_{x}$,
$E_{y}$ and $E_{z}$ indeed form a POVM in the space of physical
states, owing to Eq.~(\ref{xyz}). The above derivation was,
admittedly, a rather circuitous route for obtaining a POVM for
polarization. This is due to the fact that the latter is a
secondary variable, subject to super\-selection rules.
Unfortunately, this is the generic situation.

The entire effective density matrix is reconstructed using
techniques of Chuang and Nielsen (1997), and we get a simple
expression for the reduced density matrix corresponding to the
polarization state $|\bal(\bk)\9$:
\beq
\rho_{mn} =\int\!\!
 d\mu(\bk)|f(\bk)|^2\6\bal(\bk)|\bb_m(\bk)\9\6\bb_n(\bk)|\bal(\bk)\9
 \label{reduced}
\eeq
It is interesting to note that this derivation gives a direct
physical meaning to the naive definition of a reduced density
matrix,
\beq
\rho^{\rm naive}_{mn}=\int d\mu(\bk)|f(k)|^2\bal_m(\bk)\bal_n^*(\bk)
=\rho_{mn}.
\eeq
Since polarization 3-vectors transform under rotations regardless
of momentum, the effective $3\times 3$ polarization density matrix
has a standard transformation law under rotation $R$ as well,
$\rho\rightarrow R\rho R^T$.

Our basis states $|\bk,\bep_\bk\9$ are direct products of momentum
and polarization. Owing to the transversality requirement
$\bep_\bk\cdot\bk=0$, they remain direct products under Lorentz
transformations. All the other states have their polarization and
momentum degrees of freedom entangled. As a result, if one is
restricted to polarization measurements as described by the POVM
elements (\ref{povmexa}), {\it there do not exist two orthogonal
polarization states\/}. It follows that photon polarization states
cannot be cloned perfectly, because the no-cloning theorem
(Wootters and Zurek, 1982; Dieks, 1982) forbids an exact copying
of unknown non-orthogonal states. In general, any measurement
procedure with finite momentum sensitivity will lead to the errors
in identification.

Our present problem is the distinguishability by our observer,
Bob, of a pair of different quantum states that were prepared by
Alice. The probability of an error by Bob is  given by
Eq.~(\ref{pe}). The distinguishability of polarization density
matrices depends on the observer's motion. We again assume that
Bob moves along the $z$-axis with a velocity $v$.  Let us
calculate his reduced density matrix. Recall that reduced density
matrices have no transformation law (only the {\it complete\/}
density matrix has one) except in the limiting case of sharp
momenta.  To calculate Bob's reduced density matrix, we must
transform the complete state, and only then take a partial trace.
A detailed calculation (Peres and Terno, 2003) leads to
\beq
P'_E=\frac{1+v}{1-v}P_E, \label{dope}
\eeq
which may be either larger or smaller than $P_E$. As expected, we
obtain for one-photon states the same Doppler effect as in the
classical equation (\ref{doppler}).

\subsection{Entanglement}

An important problem is the relativistic nature of quantum
entanglement when there are several particles. For two particles,
an invariant definition of the entanglement of their spins would
be to compute it in the Lorentz ``rest frame'' where
$\6\sum\pp\9=0$. However, this simple definition is not adequate
when there are more than two particles, because there appears a
problem of cluster decomposition: each subset of particles may
have a different rest frame. This is a difficult problem, still
awaiting for a solution. We shall mention only a few partial
results.

First, we have to define a convenient measure of entanglement. For
two spin-\half\ particles, the {\it concurrence\/}, $C(\rho)$, is
defined as follows (Wootters, 1998). Introduce a spin-flipped
state
$\tilde{\rho}=(\sigma_y\0\sigma_y)\rho^*(\sigma_y\0\sigma_y)$. The
concurrence is
\beq
C(\rho)=\max(0,\,\lambda_1-\lambda_2-\lambda_3-\lambda_4),
\eeq
where $\lambda_i$ are the eigenvalues, in decreasing order, of the
Hermitian matrix $[\sqrt{\rho}\tilde{\rho}\sqrt{\rho}]^{1/2}$. The
larger the concurrence, the stronger the entanglement: for
maximally entangled states $C=1$, while for non-entangled states
$C=0$.

Alsing and Milburn (2002) considered bipartite states with
well-defined momenta.  They showed that while Lorentz
transformations change the appearance of the state in different
inertial frames and the spin directions are Wigner rotated, the
amount of entanglement remains intact. The  reason is that Lorentz
boosts do not create spin-momentum entanglement when acting on
eigenstates of momentum, and the effect of a boost on a pair is
implemented on both particles by local unitary transformations,
which are known to preserve entanglement.  The same conclusion is
valid for photon pairs.

In particular, Hacyan (2001) showed that since the polarization
angle remains constant in the polarization plane, the directions
of perfect correlation for two photons still exist in any
reference frame, even if they are different from the laboratory
directions. Terashima and Ueda (2003) showed that in a quite
general setting for both massive and massless particles, allowing
for relative motion, it is always possible to find directions of
perfect (anti)correlations.

However, realistic situations involve wave packets. For example, a
state of two spin-$\half$ particles is
\beq
|\Upsilon_{12}\9=\!\sum_{\sigma_1,\sigma_2}\int\!
d\mu(p_1)d\mu(p_2)g(\sigma_1\sigma_2,\bp_1,\bp_2)|\bp_1,\sigma_1;
\bp_2,\sigma_2\9,
 \label{entfig}
\eeq
where $d\mu(p)=d^3{\bf p}/16\pi^3p^0$ as usual.

For typical particle beams, $g$ is sharply peaked at some values
$\bp_{10}$, $\bp_{20}$. Again, a boost to any Lorentz frame will
result in a unitary $U(\Lambda)\otimes U(\Lambda)$ acting on each
particle separately, thus preserving the entanglement.
Nevertheless, since boosts can change entanglement between
different degrees of freedom of each particle, the spin-spin
entanglement is frame-dependent as well.

Gingrich and Adami (2002) investigated the reduced density matrix
for $|\Upsilon_{12}\9$ and made explicit calculations for the case
where $g$ is a Gaussian, as in the work of Peres, Scudo, and Terno
(2002). They showed that if  two particles are maximally entangled
in a common, approximate rest frame (Alice's frame), then
$C(\rho)$, as seen by a Lorentz-boosted Bob, decreases when the
boost velocity tends to $c$. Of course, the inverse transformation
from Bob to Alice will increase the concurrence.  Thus, we see
that that spin-spin entanglement is not a Lorentz invariant
quantity, exactly as spin entropy is not a Lorentz scalar.
Relativistic properties of the polarization entanglement we
investigated by Bergou, Gingrich and Adami (2003).

\subsection{Communication channels}\label{cocha}

Although reduced polarization density matrices have no general
transformation rule, the above results show that such rules can be
established for particular classes of experimental procedures. We
can then ask how these effective transformation rules,
$\tau'=T(\tau)$, fit into the framework of general state
transformations. Are they completely positive (CP) as in
Eq.~(\ref{ArhoA})?  It can be proved that distinguishability, as
expressed by natural measures like $P_E$, cannot be improved by
any CP transformation (Fuchs and van de Graaf, 1999). However, the
CP requirement may fail if there is a prior entanglement with
another system and the  dynamics is not factorizable (Pechukas,
1994; \v Stelmachovi\v c and Bu\v zek, 2001; Salgado and
S\'anchez-G\'omez, 2002).

Since in Eq.~(\ref{dope}) and in the discussion following
Eq.~(\ref{pe}) we have seen that distinguishability {\it can\/} be
improved, we conclude that these transformations are {\it not
completely positive\/}. The reason is that the Lorentz
transformation acts not only on the ``interesting'' discrete
variables, but also on the primary momentum variables that we
elected to ignore and to trace out, and its action on the
interesting degrees of freedom depends on the ``hidden'' primary
ones. Of course, the complete state, with all the variables,
transforms unitarily and distinguishability is preserved.

This technicality has one important consequence. In quantum
information theory quantum channels are described by completely
positive maps that act on qubit states (Holevo, 1999; Keyl, 2002).
Qubits themselves are realized as discrete degrees of freedom of
various particles. If relativistic motion is important, then not
only does the vacuum behave as a noisy quantum channel, but the
very representation of a channel by a CP map fails.

\section{The role of quantum field theory}

The POVM formalism is an essential tool of quantum information
theory. Entanglement is a major resource for quantum communication
and computation. In this section we present results of quantum
field theory that are important for the relativistic
generalization of these concepts. Mathematical results are stated
in an informal way. Rigorous formulations and fine mathematical
points can be found in the references that are supplied for each
concept or theorem we introduce.

\subsection{General theorems}

First, we define the notions of local and quasi-local operators
(Emch, 1972; Bogolubov \etal, 1990; Haag, 1996; Araki, 1999).
Local operators are associated with bounded regions of spacetime.
For example, they may be field operators that are smeared with
functions of bounded support (that is, functions that vanish if
their argument is outside of a prescribed bounded region $\cO$ of
spacetime).  Smeared renormalized stress-energy tensors also
belong to this category. Quasi-local operators are obtained when
the smearing functions have exponentially decaying tails.\medskip

\noindent {\bf Theorem.} The set of states $\cA(\cO)|\Omega\9$,
generated from the vacuum $|\Omega\9$ by the (polynomial) algebra
of operators in any bounded region, is dense in the Hilbert space
of all field states. \hfill $\Box$

\medskip This is the Reeh-Schlieder theorem (Reeh and Schlieder, 1961;
Streater and Wightman, 1964; Haag 1996; Araki, 1999).  It asserts
that there are local operators $Q\in\cA(\cO)$ which, applied to
the vacuum, produce a state which is arbitrarily close to any
arbitrary $|\Upsilon\9$ (the vacuum state can be replaced by any
state of finite energy).  Thus in principle any entangled state
can be arbitrarily closely approximated by suitable local
operations on any other state.

The theorem reveals a surprising amount of entanglement that is
present in the vacuum state $|\Omega\9$. The corollary below shows
that if a local operator is used to model a detector, that
detector must have ``dark counts'': it has a finite probability to
``click'' in a vacuum.\medskip

\noindent {\bf Corollary.} No operator that is localized in a bounded
spacetime region annihilates the vacuum (nor any other physical
state).  \hfill $\Box$

\medskip Another important theorem is due to Epstein, Glaser and Jaffe
(1965):\medskip

\noindent {\bf Theorem.} If a field $Q(x)$ satisfies
$\6\Psi|Q(x)|\Psi\9\geq0$ for all states, and if
$\6\Omega|Q(x)|\Omega\9=0$ for the vacuum state, then $Q(x)=0$.
\hfill $\Box$

\medskip This implies that no POVM
constructed from local or quasi-local operators can have zero
vacuum response. The theorem predicts for any local field $Q(x)$
that has a zero vacuum expectation value, namely
$\6\Omega|Q(x)|\Omega\9=0$, there exists a state for which the
expectation value of $Q(x)$ is negative. Further details can be
found in the original article and in Tippler (1978).

Another implication is a violation of the classical energy
conditions (Hawking and Ellis, 1973; Wald, 1984). Classically,
energy density is always positive and the stress-energy tensor for
all classical fields satisfies the weak energy condition (WEC)
$T_{\mu\nu}u^\mu u^\nu\geq 0$, where $u^\mu$ is any timelike or
null vector. The Epstein-Glaser-Jaffe theorem shows that this is
impossible for the renormalized stress-energy tensor of quantum
field theories. Since it has by definition a null vacuum
expectation value, there are states $|\Upsilon\9$ such that
$\6\Upsilon|{T}_{\mu\nu}u^\mu u^\nu|\Upsilon\9 < 0$. For example,
squeezed states of the electromagnetic field (Mandel and Wolf,
1995), or the scalar field (Borde, Ford, and Roman, 2002), have
locally negative energy densities. The violation of WEC raises
doubts on the use of energy density for the description of
particle localization, as discussed in Sec.~\ref{loc}.

While any entangled state can be approximated by the action of
local operators on $|\Omega\9$, the clustering  property  of the
vacuum\footnote{Its relation to the cluster property of the
$S$-matrix is discussed by Weinberg (1995).} asserts that states
created by local operations, namely $Q|\Omega\9,\; Q\in\cA(\cO)$,
tend to look practically like a vacuum with respect to
measurements in distant, causally unconnected regions. The
behavior of detectors that are far away from each other is ruled
by the following theorems, where, for a local operator
$B\in\cA(\cO)$, we denote by $B_\bx$ its translate by a spatial
vector \bx, {\it i.e.\/}, $B_\bx= U(\bx)BU^\dag(\bx)$.\medskip

\noindent {\bf Theorem.} If  $A, B\in \cA(\cO)$ are local operators
and $|\Omega\9$ is the vacuum state, then
\beq
\6\Omega|AB_\bx|\Omega\9\stackrel{{|\bx|\to\infty}}{\longrightarrow}
\6\Omega|A|\Omega\9 \6\Omega|B_\bx|\Omega\9. \label{cluster}
\eeq

\medskip There are estimates on the rate of convergence of the above
expression as a function of the spacelike separation for the cases
of massive and massless particles. The asymptotic behavior depends
on that of the Wightman function $W(x_1,x_2)$ for
$|x_1-x_2|^2\to\infty$ (Streater and Wightman, 1964; Bogolubov
\etal\ 1990; Haag, 1996).\medskip

\noindent {\bf Theorem.} If $A\in\cA(\cO_1)$ and $B\in\cA(\cO_2)$,
where $\cO_1$ and $\cO_2$ are mutually spcelike regions with a
spacelike separation $r$, then
\beq
|\6\Omega|AB|\Omega\9-\6\Omega|A|\Omega\9\6\Omega|B|\Omega\9|
\eeq
for a {\it massless\/} theory is bounded by
\beq
f(\cO_1,\cO_2,A,B)/r^2,\label{photonvac}
\eeq
where $f$ is a certain function that depends on the regions and
the operators, but not on the distance between the regions; for a
{\it massive\/} theory it is bounded by
\beq
e^{-mr}g(A,B),
\eeq
where $m$ is the relevant mass and $g$ depends on the operators
only. In this case $\cO_1,\cO_2$ may be unbounded.\hfill $\Box$

\medskip The explicit derivation of the coefficients requires a more
detailed treatment. Particular cases and values of numerical
constants are given by Emch (1972), Fredenhagen (1985), Haag
(1996), and Araki (1999).

While it seems that vacuum correlations for massless fields decay
much slower, the difference disappears if the finite sensitivity
of detectors for soft photons is taken into account. It was shown
by Summers and Werner (1987) that if a detector has an energy
threshold $\epsilon$, the latter serves as an effective mass in
correlation estimates, and an additional $e^{-\epsilon r}$ factor
appears in Eq.~(\ref{photonvac}).

\subsection{Particles and localization}\label{loc}

Classical interventions in quantum systems are localized in space
and time. However, the principles of quantum mechanics and
relativity dictate that this localization is only approximate. The
notion of particles has an operational meaning only owing to their
localization: particles are what is registered by detectors.

When quantum mechanics was a new science, most physicists wanted
to preserve the notions with which they were familiar, and
considered particles as real objects having positions and momenta
that were possibly unknown, and/or subject to an ``uncertainty
principle.'' Still, a few writers expressed critical opinions, for
example ``\ldots no scheme of operations can determine
experimentally whether physical quantities such as position and
momentum exist\ldots\ we get into a maze of contradictions as soon
as we inject into quantum mechanics such concepts carried over
from the language of our ancestors\ldots'' (Kemble, 1937).

More recently, Haag (1996) wrote

\begin{quote} ``\ldots it is not possible to assume that an electron
has, at a particular instant of time, any position in space; in
other words, the concept of position at a given time is not a
meaningful attribute of the electron. Rather, `position' is an
attribute of the interaction between the electron and a suitable
detection device.''
\end{quote}

We shall first briefly examine some aspects of the old fashioned
approach to localization.  First we note that even when we
construct a local probability density (and, possibly, a
corresponding current) it is impossible to interpret
$\rho(\bx,t)d^3\bx$ as the probability to find a particle in the
volume $d^3\bx$ at the space point $\bx$. It was argued by Landau
and Peierls (1931) that a particle may be localized only with
uncertainty $\Delta x>\hbar c/\6E\9$, where $\6E\9$ is the
particle's expected energy. Intuitively, confinement of a particle
to a narrower domain by ``high walls'' requires a very strong
interaction which leads to pair production. Haag and Swieca (1965)
have shown that restriction to a compact region of spacetime makes
it impossible to detect with certainty any state. Hegerfeldt
(1985) proved that if a one-particle POVM leads to probability
distributions such that  the total probability of finding a
particle outside a sphere of radius $R$ at time $t$ is bounded by
\beq
{\rm Prob}_{\not\in R}< C^2\exp(-2\gamma R),
\eeq
where $C$ is some constant and $\gamma>m$, then at later times the
probability distribution will spread faster than light.
Furthermore, Giannitrapani (1998) and Toller (1999) proved that a
spacetime localized POVM cannot be constructed even from
quasi-local operators. General discussions of localization from
the point of view of algebraic quantum field theory can be found
in the works of Buchholz and Fredenhagen (1982), Roberts (1982),
Neumann and Werner (1983), Werner (1986) and Haag (1996).

Much earlier, Newton and Wigner (1949) had attempted to define a
position operator, whose spectral decomposition (Wightman, 1962)
gives a rough indication of the particle localization. However, it
was shown by Rosenstein and Usher (1987) that Gaussian-like
Newton-Wigner wave functions lead to superluminal propagation of
probability distributions. Busch (1999) reviewed the problems
involved in the construction of POVMs for particle localization.

Energy density is directly related to photon localization in
quantum optics (Mandel and Wolf, 1995; Bialynicki-Birula, 1996).
If the electrons in a detector interact with the electric field of
light, then in a simple model the detection probability is
proportional to the expectation value of the normal-ordered
electric field intensity operator $I(\bx,t)$ (Mandel, 1966), and
the latter is proportional to the energy density. This probability
distribution decays asymptotically as the seventh power of
distance, or even slower (Amrein, 1969). Despite it success in
these examples, the notion of localization based on the energy
density cannot have a universal validity, because the violation of
WEC makes it unsuitable for the construction of POVMs.

The real physical problem is how localized {\it detectors\/} can
be. The idealization of ``one detector per spacetime point'' is
obviously impossible. How can we manage to ensure that two
detectors have zero probability to overlap? There appears to be a
fundamental trade-off between detector reliability and
localizability. The bottom line is how to formulate a relativistic
interaction between a detector and the detected system. A true
detector should be amenable to a dual quantum-classical
description, as in the Hay-Peres model (1998). This problem seems
to be very far from a solution. Completely new notions may have to
be invented.

Although states with a definite number of particles are a useful
theoretical concept, a look at quantum optics techniques or at the
{\it Table of Particle Properties\/} shows that experimentally
accesible quantum states are usually not eigenstates of particle
number operators. In general any process that is not explicitly
forbidden by some conservation law has a non-zero amplitude
(Weinberg, 1995; Peskin and  Schroeder, 1995; Haag, 1996). There
are multiple decay channels, extra soft photons may always appear,
so that the so-called `one-photon' states are often accompanied by
soft multiphoton components,
\beq
\alpha|\Omega\9+\beta|1_\omega\9+\gamma|2_{\omega'\omega''}\9+\ldots,
\qquad |\beta|\sim 1.
\eeq
Thus the physical realization of a single qubit is itself
necessarily an idealization.

\subsection{Entanglement in quantum field theory}

Recall that while the Reeh-Schlieder theorem ensures that any
state can be approximated by local operations, the clustering
property of the vacuum implies that locally created states look
almost like a vacuum for distant measurements. The Reeh-Schlieder
and Epstein-Glaser-Jaffe theorems entail dark counts for local
detectors. The responses of spatially separated detectors are
correlated, but these correlations decay fast due to cluster
properties.

We now consider correlation experiments with devices $a$ and $b$
placed in spacelike-separated regions $\cO_L$ and $\cO_R$, so all
local operators pertaining to these regions commute:
$[\cA(\cO_L),\cA(\cO_R)]=0$. In each region, there are two such
devices, labelled $a_1,a_2,b_1,b_2$, which yield outcomes ``yes"
or ``no" in each individual experiment. We denote the
probabilities for positive outcomes as $p(a_j)$ and $p(b_k)$, and
by $p(a_j\wedge b_k)$ the probability of their joint occurrence.

The measuring apparatus $a_j$ is described by a POVM element
$F_j\in\cA(\cO_L)$ and the probability of the ``yes" outcome for a
state $\rho$ is $\tr(\rho F_j)$. If $G_k$ is the POVM for
apparatus $b_k$ then the probability of the ``yes-yes" outcome is
$\tr(\rho\,F_j G_k)$. Let us to introduce operators $A_j=2F_j-\1$
and $B_k=2G_k-\1$, and define
\beq
\zeta(a,b,\rho)=\half\tr \{\rho[A_1(B_1+B_2)+A_2(B_1-B_2)]\}.
\eeq
This quantity, which is experimentally measurable, has a classical
analogue whose value is bounded: $\zeta\leq1$. This is the CHSH
inequality (Clauser \etal, 1969), which is one of the variants of
the Bell inequality (Bell, 1964).\footnote{Recall that the Bell
inequalities are essentially classical (Peres, 1993). Their
violation by a quatum system is a sufficient condition for
entanglement, but not a necessary one.}

The above definition of $\zeta$ can be extended to
\beq
\zeta(\cA,\cB,\rho)=\sup\zeta(a,b,\rho),
\eeq
where $\cA=\cA(\cO_L)$, $\cB=\cA(\cO_R)$, and the supremum is
taken over all operators $A_j,B_k$.  It was shown by Cirel'son
(1980) that there is also a quantum bound on correlations: for
commuting algebras $\cA$ and $\cB$ and any state $\rho$,
\beq
\zeta(\cA,\cB,\rho)\leq\sqrt{2}.
\eeq

Further results of Summers and Werner (1985, 1987a,b) and Landau
(1987) establish that a violation Bell's inequalities is generic
in quantum field theory. For any two spacelike separated regions
and any pairs of operators, $a$, $b$, there is a state $\rho$ such
that the CHSH inequality is violated, namely, $\zeta(a,b,\rho)>1$.
With additional technical assumptions the existence of a maximally
violating state $\rho_m$ can be proved:
\beq
\zeta(a,b,\rho_m)=\sqrt{2} \label{violall},
\eeq
for any spacelike separated regions $\cO_L$  and $\cO_R$. It
follows from convexity arguments that states that maximally
violate Bell inequalities are pure. What are then the operators
that lead to maximal violation? Summers and Werner (1987a) have
shown that operators $A_j$ and $B_k$ that give $\zeta=\sqrt{2}$
satisfy $A_j^2=\1$ and $A_1A_2+A_2A_1=0$, and likewise for $B_k$.
If we define $A_3:=-i[A_1,A_2]/2$, then these three operators have
the same algebra as Pauli spin matrices (Summers,1990).  Even if
we ignore the problem of localization (Sec.~V.B), a violation of
Bell inequalities  is not at all trivial, as the analysis of
various relativistic spin operators shows (Terno, 2003). For
example, for moving observers, if the observables are constructed
by means of the Pauli-Lubanski operator, the amount of violation
of Bell's inequality decreases with increasing velocity, and the
inequality is satisfied in the ultra-relativistic limit (Czachor,
1997; Ahn \etal\/, 2003).

The violation of Bell's inequalities by the vacuum state does not
mean that it is enough to have two detectors and check their dark
count coincidences. The cluster theorem predicts a strong damping
of the violations with distance.  When the lowest relevant mass is
$m>0$, clustering leads to the estimate
\beq
\zeta(\cA(\cO_L),\cA(\cO_R),\Omega)\leq1+4\exp[-mr(\cO_L, \cO_R)],
\eeq
where $r(\cO_L,\cO_R)$ is the separation between the regions
(Summers and Werner, 1985, 1987a,b). For massless particles, the
energy threshold for photodetection serves as an effective mass.
Therefore, a direct observation of vacuum entanglement should be
extremely difficult. Reznik (2000) proposed a method to convert
vacuum entanglement into conventional bipartite entanglement. It
requires to switch on and off in a controllable way  the
interaction between two-level systems and a field. Appropriately
tailored local interaction Hamiltonians can then transfer vacuum
entanglement to atoms.

The classification of entangled states and their manipulation are
current research topics in quantum information theory. Up to now
we have dealt with entanglement of a finite number of degrees of
freedom, or spin-momentum entanglement. After introducing Lorentz
transformations, we were still able to use the standard techniques
of the non-relativistic theory.  However, in the general case,
infinite-dimensional Hilbert spaces are involved. Recently Parker,
Bose, and Plenio (2000), Eisert, Simon, and Plenio (2002), and
Keyl, Schlingemann, and Werner (2003) investigated the
entanglements of formation and of distillation in
infinite-dimensional systems.

When the Hilbert space of a bipartite system is infinite
dimensional, some peculiarities arise. For pure states, a natural
measure of entanglement is the von Neumann entropy
$S=-\tr\rho\ln\rho$ of either one of the reduced density matrices.
It can be shown (Eisert, Simon and Plenio, 2002) that in an
arbitrarily small neighborhood of any state there is an infinity
of entangled states. The reason is that in the neighborhood of any
state with finite energy, there are states of infinite entropy
(Wehrl, 1978).\footnote{The set of states with infinite entropy is
trace-norm dense in the state space.} This seems paradoxical, but
if we consider states with bounded energy only, the continuity of
the degree of entanglement is restored.

Keyl, Schlingemann and Werner (2003) applied techniques of
operator algebra to systems with an infinite number of degrees of
freedom. A usuful device in the description of infinite sytems is
the notion of singular states, which cannot be represented by
density operators: states are considered to be just positive
linear functionals on the space of POVMs, and only non-singular
states are represented by density operators (Emch, 1972; Bratteli
and Robinson, 1987). One of their results is a rigorous
description to the original EPR (1935) state, which can be modeled
as a sequence of more and more squeezed two-mode states, and
actually is a singular state.

Pachos and Solano (2003) discussed the generation of entangled
states and performed {\it ab initio\/} QED calculations for the
case of two interacting spin-\half\ charged particles. They
obtained particular results for low energy scattering, and more
general situations are under  investigation.

\subsection{Accelerated detectors}\label{unruh}

In quantum field theory, the vacuum is defined as the lowest
energy state of a field. A free field with linear equations of
motion can be resolved into normal modes, such as standing waves.
Each mode has a fixed frequency and behaves as a harmonic
oscillator. The zero point motion of all these harmonic
oscillators is called ``vacuum fluctuations''and the latter, under
suitable conditions, may excite a localized detector that follows
a trajectory $x^\nu(\tau)$ parametrized by its proper time $\tau$.
The internal structure of the detector is described by
non-relativistic quantum mechanics, so that we can indeed assume
that it is approximately localized, and it has discrete energy
levels $E_n$.  Furthermore, we assume the existence of a linear
coupling of an internal degree of freedom, $\mu$, of the detector,
with the scalar field $\phi(x(\tau))$ at the position of the
detector. First-order perturbation theory gives the following
expression for the transition probability per unit proper time:
\beq
g^2\sum_n|\6E_n|\mu|E_0\9|^2\int d\tau e^{-i(E-E_0)\tau}W(\tau),
 \label{autocorrel}
\eeq
where $g$ is a coupling constant and
\beq
W(\tau)\equiv W(x(\tau_1),x(\tau_2)),\qquad \tau=\tau_1-\tau_2,
\eeq
is the Wightman function, defined by
$W(x_1,x_2)=\6\Omega|\psi(x_1)\psi(x_2)|\Omega\9$ for two
arbitrary points on the detector's trajectory (Streater and
Wightman, 1964). The integral in Eq.~(\ref{autocorrel}) is the
Fourier transform of the autocorrelation.  In other words, it
gives the power spectrum of the Wightman function.

For inertial detectors (that is, $x^\nu=v^\nu\tau$ with a constant
four-velocity $v^\nu$) the transition probability is zero, as one
should expect. However, the response rate does not vanish for more
complicated trajectories. Consider in particular one with constant
proper acceleration $a$. With an appropriate choice of initial
conditions, it corresponds to the hyperbola $t^2+x^2=1/a^2$, shown
in Fig.~3. Then the transition rate between levels appears to be
the same as for an inertial detector in equilibrium with thermal
radiation at temperature $T=\hbar a/2\pi c\kB$. This phenomenon is
called the Unruh (1976) effect. It was also discussed by Davies
(1975) and it is related to the fluctuation-dissipation theorem
(Candelas and Sciama, 1977) and to the Hawking effect that will be
dicussed in the next section.\footnote{Properties of detectors
undergoing circular acceleration, as in high energy accelerators,
were investigated by Bell and Leinaas (1983), Levin, Peleg, and
Peres (1993), and by Davies, Dray, and Manogue (1996).} A rigorous
proof of the Unruh effect in Minkowski spacetime was given by
Bisognano and Wichmann (1976) in the context of axiomatic quantum
field theory, thus establishing that the Unruh effect is not
limited to free field theory.

\begin{figure}[htbp]
\epsfxsize=0.39\textwidth
\centerline{\epsffile{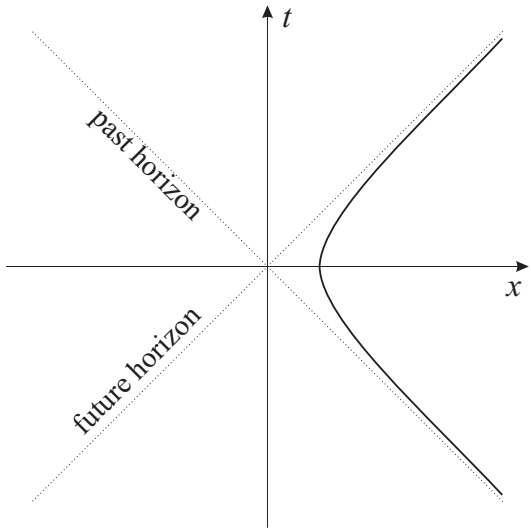}} \vspace*{-0.1cm} \caption{\small{
Dependence of the spin entropy $S$, in Bob's frame, on the values
of the angle $\theta$ and a parameter
$\Gamma\approx[1-(1-\beta^2)^{1/2}]\Delta/m\beta$, where $\Delta$
is the momentum spread in Alice's frame.}}
\end{figure}

For any reasonable acceleration, the Unruh temperature is
incomparably smaller that the black-body temperature of the cosmic
background, or any temperature ever attained in a laboratory, and
is not observable. Levin, Peleg, and Peres (1992) considered the
effect of shielding a hypothetical experiment from any parasitic
sources. This, however, creates a radically new situation, because
the presence of a boundary affects the dynamical properties of the
quantum field by altering the frequencies of its normal modes.
Finite-size effects on fields have been known for a long time,
both theoretically (Casimir, 1948) and experimentally (Spaarnay,
1958).  Levin, Peleg, and Peres showed that if the detector is
accelerated together with the cavity that shields it, it will {\it
not\/} be excited by the vacuum fluctuations of the field. On the
other hand, an inertial detector freely falling within such an
accelerated cavity will be excited. The relevant property in all
these cases is the {\it relative acceleration of the detector and
the field normal modes.}

We now consider the evolution of an arbitrary quantum system. An
observer at rest (Alice) can describe the quantum evolution on
consecutive parallel slices of spacetime, $t={\rm const.}\,$ What
can Bob, the accelerated observer, do? From Fig.~3, one sees that
there is no communication whatsoever between him and the region of
spacetime that lies beyond both horizons. Where Alice sees a pure
state, Bob has only a mixed state. Some information is lost. We
shall return to this subject in the next, final section.

\section{Beyond special relativity}
It took Einstein more than ten years of intensive work to progress
from special relativity to general relativity. Despite its name,
the latter is not a generalization of the special theory, but a
radically different construct: spacetime is not only a passive
arena where dynamical processes take place, but has itself a
dynamical nature. At this time, there is no satisfactory quantum
theory of gravitation (after seventy years of efforts by leading
theoretical physicists).

In the present review on quantum information theory, we shall not
attempt to use the full machinery of general relativity, with
Einstein's equations.\footnote{Concepts of quantum information
were recently invoked in several problems of quantum gravity and
quantum cosmology, but we restrict ourselves to conventional black
hole physics.} We still consider spacetime as a passive arena,
endowed with a Riemannian metric, instead of the Minkowski metric
of special relativity. The difference between them is essential:
it is necessary to introduce notions of topology, because it may
be impossible to find a single coordinate system that covers all
of spacetime. To achieve that result, it may be necessary to use
several coordinate patches, sewed to each other at their
boundaries. Then in each patch, the metric is not geodesically
complete: a geodesic line stops after a finite length, although
there is no singularity there. The presence of singularities
(points of infinite curvature) is another consequence of
Einstein's equations. It is likely that these equations, which
were derived and tested for the case of moderate curvature, are no
longer valid under such extreme conditions. We shall not speculate
on this issue, and we shall restrict our attention to the behavior
of quantum systems in the presence of {\it horizons\/}, in
particular of black holes. Before we examine the latter, let us
first return to entanglement, now in curved spacetime, and to the
Unruh effect, still in flat spacetime, but described now in an
accelerated coordinate system.

\subsection{Entanglement revisited}\label{unruh2}

Calculations on EPRB correlations require a common reference
frame. Only then can statements such as ``if $m_{1z}$=\half, then
$m_{2z}$=--\half'' have an operational meaning. In a curved space
we can choose an arbitrary frame at one spacetime point and then
translate it parallel to itself along a geodesic path. For
example, spin-\half\ particles may be sent to Alice and Bob, far
away.  After a reference frame is chosen at the emission point,
local frames are established for them by parallel transport along
the particles' trajectories. However, particles only approximately
follow classical geodesic trajectories, and this inevitably
introduces uncertainties in the definition of directions. Using
path integral methods, von Borzeszkowski and Mensky (2000) have
shown that if certain conditions are met, approximate EPR
correlations still exist, but ``the longer the propagation and the
stronger the gravitational field, the poorer is the correlation''.

One of the difficulties of quantum field theory in curved
spacetimes is the absence of a unique (or preferred) Hilbert
space, the reason being that different representations of
canonical commutation or anticommutation relations lead to
unitarily inequivalent representations (Emch, 1972; Bogolubov
\etal, 1990; Haag 1996). For the Minkowski spacetime, the existence of
a preferred vacuum state enables us to define a unique Hilbert
space representation. A similar construction is also possible in
stationary curved spacetimes (Fulling, 1989; Wald, 1994). However,
in a general globally hyperbolic spacetime this is impossible, and
one is faced with multiple inequivalent representations.

Genuinely different Hilbert spaces with different density
operators and POVMs apparently lead to predictions that depend on
the specific choice of the method of calculation. The algebraic
approach to field theory can resolve this difficulty for PVMs. The
essential ingredient is the notion of physical equivalence (Emch,
1972; Araki, 1999; Wald, 1994), which allows to extend the
formalism of POVMs and CP maps to general globally hyperbolic
spacetimes (Terno, 2002).

The simplest example of inequivalent representations occurs in the
discussion of the Unruh effect, when we wish to use quantum field
theory in the Rindler wedge $x>|t|$ where the detector moves, or
in the opposite wedge $x<-|t|$, which is causally separated from
it, or in both wedges together.  Each one of the two wedges, or
both together, can be considered as spacetimes on their own right
(Rindler spaces), where a global timelike field is obtained from
the set of all hyperbolas with different values of the
acceleration (Wald, 1984).

The transformation between Minkowski and Rindler wedge
descriptions are unitary only formally (Unruh and Wald 1984; Wald
1994) and algebraic field theory should be used to give a rigorous
interpretation to these formal expressions (Emch, 1972; Haag,
1996). A quantum field theory can be defined in a standard way
because the Rindler spaces are globally hyperbolic. They admit a
Cauchy surface for specifying initial values, whose domain of
development is the entire spacetime (Hawking and Ellis, 1973;
Wald, 1984 and 1994).  The vacuum state $|0_R\9$ obtained in this
construction is called a Rindler vacuum. It is a natural vacuum
for observers who move on orbits like in Fig.~3, with different
positive values of the acceleration $a$.

As a consequence of the Reeh-Schlieder theorem, it follows that a
Minkowski vacuum $|\Omega\9$ corresponds to a mixed state in the
Rindler spacetime. To relate the Minkowski and Rindler Hilbert
spaces, fields in both wedges are required.  The relation between
the standard Minkowski Fock space and a tensor product of Rindler
Fock spaces is given by a formally unitary operator $U$, whose
action on the Minkowski vacuum is
\beq
U|\Omega\9=\prod_i\sum_{n=0}^\infty
\exp(-n\pi\omega_i/a)|n_{iL}\9\otimes|n_{iR}\9,
\eeq
where $\omega_i$ denotes the frequencies of the modes of the
Rindler fields, and $n_i$ are the corresponding occupation
numbers. The above expression suggests that the Minkowski vacuum
has the structure of a maximally entangled state when viewed by
accelerated observers. When restricted to only one wedge, the
state becomes
\beq
\rho=\prod_i\sum_{n=0}^\infty\exp(-n\pi\omega_i/a)Z^{-1}_i|n_{iR}\9\6n_{iR}|,
\eeq
where the {\it i\/}th mode was normalized by $Z_i=\sum_i
\exp(-n\pi\omega_i/a)$. That state indeed produces a thermal density
matrix $\rho\propto\exp(-H_R/T)$, where $H_R$ is the field
Hamiltonian for region $R$, and $T=a\hbar/2\pi c\kB$. We can now
calculate the entanglement of the Minkowski vacuum as seen by an
accelerated observer. A natural reduced density matrix is $\rho$
itself, which is a singular state (in the sense of Sec.~V.C) of an
infinite thermal bath. Its entropy is infinite, which is in
agrement with the previous discussion, since the energy of such a
system is also infinite.

The relationship between Minkowski and Rindler wave packets was
analyzed by Audretsch and M\"{u}ller (1994a). These authors also
discussed local detection by Rindler observers and EPR-like
correlations (Audretsch and M\"{u}ller, 1994a, b).

Alsing and Milburn (2002, 2003) examined the fidelity of
teleportation from Alice in an inertial frame to Bob who is
uniformly accelerated. Assume that qubits are realized by some
mode $\omega$ of the electromagnetic field, and that Alice's state
is $|\Psi\9=\alpha|\Omega\9+\beta|1\9$, where $|\Omega\9$ is the
Minkowski vacuum. Then the best state that Bob can hope to get is
\beq
|\Psi'\9=\alpha|0_R\9+\beta|1_R\9,
\eeq
where $|0_R\9$ is the Rindler vacuum, and some mode $\omega'$ (as
seen by Bob) was chosen for his realization of qubits. The
fidelity of teleportation $|\Psi\9 \to |\Psi'\9$ then decreases
with Bob's acceleration. It also depends on time: the fidelity of
course vanishes when Alice is behind Bob's horizon.

\subsection{The thermodynamics of black holes}\label{terbla}

Black holes result from concentrations of matter so large that
their gravity pull prevents the escape of light (Michell, 1784;
Laplace, 1795). In other words, a future horizon is formed.  While
Unruh's horizons were for observers whose asymptotic speed
approaches $c$, a black hole horizon affects every observer. We
now present basic facts of black hole physics, limiting ourselves
almost exclusively to spherically-symmetric spacetimes.  The
literature on black holes is voluminous, and our sketch gives just
a glimpse of this fascinating subject. Our main sources for
classical black hole physics were Landau and Lifshitz (1975),
Hawking and Ellis (1973), Wald (1984) and Frolov and Novikov
(1998). For quantum aspects, we consulted Birrell and Davies
(1982), Wald (1994), Brout \etal\, (1995), and  Frolov and Novikov
(1998). An extensive survey of black hole thermodynamics was given
by Wald (1999, 2001).  In this section, unless otherwise stated,
$c=G=\hbar=1$.

Spacetime outside a spherically symmetric distribution of matter
(and hence outside an incipient black hole during all stages of
its collapse) is described by the \Sc\/ metric,
\beq
ds^2=(1-2M/r)dt^2-(1-2M/r)^{-1}dr^2-r^2d\Omega^2.
\eeq
The proper time of a stationary observer is
$d\tau=\sqrt{g_{tt}}dt=
\sqrt{1-2M/r}dt$, and the radial distance is
$dl=\sqrt{-g_{rr}}dr=(1-2M/r)^{-1/2}dr$. This metric has a
coordinate singularity at $r=2M$, which can be removed by a
transition to various alternative coordinate systems. As we shall
see, it is a kind of ``boundary'' of the black hole. On the other
hand, the singularity at $r=0$ is physical: the spacetime
curvature diverges there.

Spacetimes may have symmetries. If translation along a family of
curves leaves the metric invariant, the field of tangent vectors
to these curves is called a Killing field (Killing, 1892). Killing
vectors $\chi^\mu$ have many useful properties.

For example, the \Sc\ metric is invariant under time translations,
$t\to t+\tau$. The corresponding Killing vector
$\chi^\mu=(1,0,0,0)$ is timelike for $r>2M$ and spacelike for
$r<2M$. It becomes null on the horizon. The surface gravity
$\kappa$, which characterizes the strength of gravitational field
near the horizon, is defined as
\beq
\kappa=\lim(a\alpha),
\eeq
where $a$ is the norm of the proper four-acceleration of a
stationary object, and $\alpha$ is a red-shift factor. For \Sc\
black holes, $\alpha=\sqrt{g_{tt}}$ and $\kappa=1/4M$. Hawking and
Ellis (1973), and Wald (1984, 1999) describe many properties of
the surface gravity.  Bardeen, Carter, and Hawking (1973), have
shown that $\kappa$ is constant over the horizon of any stationary
black hole. This is known as a the zeroth law of black hole
mechanics.

Even in classical general relativity, there is a serious
difficulty with the second law of thermodynamics when a black hole
is present: if we drop ordinary matter into a black hole, it will
disappear into a spacetime singularity, together with its entropy
$S$. No compensating gain of entropy occurs, so that the total
entropy in the universe decreases.  One could attempt to salvage
the second law by invoking the bookkeeping rule that one must
continue to count the entropy of matter dropped into a black hole
as still contributing to the total entropy of the universe.
However, the second law would then be observationally
unverifiable.

It was noted by Bekenstein (1972, 1974) that properties of the
horizon area of a stationary black hole resemble those of entropy.
In the most general case, a stationary black hole is characterized
by three parameters: its mass $M$, angular momentum $J$ and charge
$Q$. The first law of black hole dynamics (Bardeen, Carter, and
Hawking, 1973; Iyer and Wald, 1994) states that
\beq
 dM=\frac{\kappa}{8\pi}dA+\Omega dJ+\Phi dQ,
\eeq
where $\Omega$ is the angular velocity and $\Phi$ the electric
potential. This relation is formally identical to the first law of
thermodynamics, if we identify temperature with surface gravity
and entropy with horizon area. We would then have
\beq
T=\frac{\kappa}{2\pi}\frac{\hbar c^3}{G\kB},\qquad
S=\frac{A}{4\lP^2}, \label{TS}
\eeq
where $\lP=\sqrt{\hbar G/c^3}$ is the {\it Planck length\/}, and
ordinary units were restored.

Bekenstein (1972, 1974)  proposed to assign to a black hole of
area $A$ an entropy
\beq S\BH=Ac^3/4\hbar G,
\eeq
thus elevating a formal analogy to the status of a physical law.
Hawking (1974) found that a black hole radiates like a black body
at temperature $T$, and thereby put the analogy between black hole
mechanics and thermodynamics on firm ground.

There are many ways to explain Hawking radiation (Hawking, 1975;
Wald, 1975; Birrell and Davies, 1982;  Fredenhagen and Haag, 1990;
Wald, 1994; Brout \etal, 1995). Here, we follow the informal
presentation of Frolov and Novikov (1998), which is based on the
analogy with pair creation by an external static field. Actually,
spacetime is not static when a star collapses into a black hole
and later evaporates. However, usually it is an excellent
approximation to treat it as static. A rigorous analysis along
these lines was made by Brout \etal\/ (1995).

Similarities between pair production and Hawking radiation were
discussed by M\"{u}ller, Greiner and Rafeski (1977).  Let $\Gamma$
be the field strength  and $g$ the charge. By analogy with the
tunnel effect, the probability that  a virtual pair of particles
be found at a distance $l$ from one another is approximately
$e^{-l/\lambda}$, where $\lambda$ is the Compton wavelength. A
pair may turn to be real if $g\Gamma l\geq 2mc^2$.  Thus, the
probability of particle creation is $w\propto\exp(-\zeta m^2
c^3/\hbar g \Gamma)$, where the numerical constant $\zeta$ can be
obtained by a more detailed calculation.

A naive application of this formula to particle creation in a
static gravitational field turns out to give not only the right
result, but also some valuable insights. In particular,
conservation of energy implies that a static gravitational field
can create particles only if there are regions with timelike
Killing fields and others with spacelike ones; a horizon is
needed. A static gravitational field without horizons cannot
create particles (Birrell and Davies, 1982; Wald, 1984). A black
hole emits particles as if it were a black body with temperature
\beq
T=\kappa/2\pi\kB,
\eeq
as in Eq.~(\ref{TS}).

The generalized second law of thermodynamics (Bekenstein, 1974;
Frolov and Page, 1993; Wald, 1994; Frolov and Novikov, 1998)
states that
\beq\Delta S+\Delta S\BH\geq 0.
\eeq
An informational analysis of this law by Hosoya, Carlini, and
Shimomura (2001) clarified its relation to classical bounds on
accessible information (Levitin, 1969, 1987; Holevo, 1973).
Bekenstein and Mayo (2001) and Bekenstein (2002) gave a
description of the information absorption and emission by black
holes in terms of quantum channels.

A natural question is what (and where) are the degrees of freedom
responsible for the black hole entropy. On this issue, there are
conflicting views. It is not clear whether we should think of
these degrees of freedom as residing outside the black hole in its
thermal atmosphere, or on the horizon in Chern-Simons states, or
inside the black hole, associated with what classically
corresponds to the singularity deep within it. Or perhaps the
microscopic origin of $S\BH$ is the entanglement between Hawking
particles inside and outside the horizon (Bombelli \etal, 1986;
Ashtekar \etal, 1994; Iorio, Lambiase, and Vitiello, 2001). It is
likely that in order to gain a better understanding of the degrees
of freedom responsible for black hole entropy, it will be
necessary to achieve a deeper understanding of the notion of
entropy itself (Zurek, 1990).

Suppose now that the matter that has fallen inside the horizon had
quantum correlations with matter that remained outside. How is
such a state described by quantum theory? Are these correlations
observable? This problem is not yet fully understood, although
such correlations play an essential role in giving to Hawking
radiation a nearly exact thermal character (Wald, 1975). It is
hard to imagine a mechanism for restoring the correlations during
the process of black hole evaporation. On the other hand, if the
correlations between the inside and the outside of a black hole
are not restored during the evaporation process, then by the time
that the black hole has evaporated completely, an initial pure
state will have evolved to a mixed state, and some ``information''
will have been lost.

Hawking's radiation resolved the thermodynamic difficulty only to
introduce another puzzle. An inevitable result of that radiation
is the evaporation of the black hole after a finite time (see
Appendix B). Since the emitted particles are overwhelmingly
massless, black hole evaporation leads to baryon number
non-conservation.

Hawking (1976, 1982) also introduced a superoperator to describe
the quantum state evolution during the  black hole formation and
evaporation (see Appendix B). A detailed analysis of this
superoperator was made by Strominger (1996). It is (at least
formally) completely positive and as such it is a perfectly normal
operation of quantum information theory (Terno, 2002).

Yet, it has often been asserted that the evolution of an initial
pure state into a final mixed state conflicts with quantum
mechanics, and this issue is usually referred to as the ``black
hole information loss paradox.''  These pessimistic views are
groundless. When black hole thermodynamics appeared in the 70's,
notions such as POVMs and completely positive maps were unknown to
the relativistic community. Today, we know that the evolution of
pure states into mixtures is the general rule when a classical
intervention is imposed on a quantum system, as we have seen in
Sec.~II. In the present case, the classical agent is the spacetime
metric itself, which is borrowed from classical general relativity
in the absence of a consistent quantum gravity theory. Attempts to
introduce a hybrid quantum-classical dynamics by using the Koopman
(1931) formalism are not mathematically inconsistent, but they
violate the correspondence principle and are physically
unacceptable (Peres and Terno, 2001). Anyway, the evolution of an
initial pure state into a final mixed state is naturally
accomodated within the framework of the algebraic approach to
quantum theory (Wald, 1994), and that of a generalized quantum
theory (Hartle, 1998).

The final fate of black holes and its relation to the information
paradox were discussed by Preskill (1993), 't Hooft (1996, 1999)
and Frolov and Novikov (1998). However, this issue may be
conclusively resolved only after there is a consistent theory of
quantum gravity, allowing meanwhile for a number of tantalizing
speculations. Here we present five of the most popular
alternatives of what happens with the ``information" when a black
hole evaporates.

\begin{itemize}

\item Information is lost: Hawking's superscattering that was
described above  is a fundamental feature of quantum theory and
not just an effective description.

\item There is no information loss: if the spectrum is analyzed
carefully, there may be enough non-thermal features to encode all
the information. Bekenstein (1993) showed that deviations of the
Hawking radiation from the black body spectrum may help
reconstruct part of the information. Hod (2002) estimated that,
under suitable assumptions about black hole quantization, the
maximal information emission rate may be sufficient to recover all
the information from the resulting discrete spectrum of the
radiation.

\item Information comes out at the end, at the Plank scale physics.
Frolov and Vilkovisky (1981) constructed a model that provides for
this possibility.

\item There is a stable black hole remnant with about the Planck mass
(0.02 $\mu$g)   and information is somehow encoded in it
(Aharonov, Casher, and Nussinov, 1987).

\item Information escapes to baby universes, that are created instead
of true singularities (Zel'dovich, 1977; Hawking, 1988). The
overall evolution of the entire multiverse is unitary, but since
baby universes are causally unconnected to our universe and the
total state is entangled, we perceive a loss of information.
\end{itemize}  \medskip

Still a different scenario is implied by the works of Gerlach
(1976) and Boulware (1976): a particle that falls into an eternal
black hole crosses the horizon after an infinite amount of the
coordinate time $t$, but only a finite amount of its own proper
time. On the other hand, the evaporation of a black hole takes a
finite amount of the coordinate time, which is the physical time
of a distant observer (see Appendix B). From the point of view of
the infalling observer, the horizon always appears to recede
before her, until it finally disappears (or shrinks to the Planck
scale) and the region ``beyond the horizon'' is unattainable. The
distant observer sees the infalling one quickly arrive arbitrarily
close to the effective horizon, then she is nearly ``frozen" there
for an exceedingly long time, and finally either the black hole
evaporates or the universe collapses.  Therefore it makes no sense
to assert that states having (essential) support on the part of
the Cauchy surface that lies beyond the horizon would be
correlated with an outgoing Hawking radiation and then
mysteriously disappear. There is no issue of information loss at
all (Sonego, Almergren, and Abramowitz, 2000; Alberghi \etal,
2001).\footnote{We have listed this opinion last, because it is
the one we tend to support.}

\subsection{Open problems}

The good news are that there is still plenty of work to be done.
Here we shall mention a few problems that appear interesting and
from which more physics can be learnt.

\begin{itemize}
\item As mentioned in Sec.~\ref{loc}, quantum field theory
implies a trade-off between the reliability of detectors and their
localization. This is an important practical problem. A proper
balance must be found between the loss of undetected signals,
false alarms (dark counts), and our knowledge of the location of
recorded events. A quantitative discussion of this problem would
be most welcome.

\item It is possible to indicate the approximate orientation of
a Cartesian frame by means of a few suitably prepared spins
(Bagan, Baig, and Mu\~noz-Tapia, 2001), or even a single hydrogen
atom (Peres and Scudo, 2001). Likewise, the quantum transmission
of the orientation of a Lorentz frame should be possible. This
problem is much more difficult, because the Lorentz group is not
compact and has no finite-dimensional unitary representations
(Wigner, 1939).

\item Progressing from special to general relativity, what is the
meaning of parallel transport of a spin? In a curved spacetime,
the result is obviously path dependent. Then what does it mean to
say that a pair of distant particles is in a singlet state? As the
rotation group O(3) is not a valid symmetry, the classification of
particles, even the usefulness of the concept of a particle,
become doubtful. Methods are known for quantization of higher spin
fields in a curved background (Birrell and Davies, 1982; Wald,
1994), but what is the operational meaning of the resulting states
and POVMs?

\item We still need a method for detection of relativistic entanglement
that involves the spacetime properties of the quantum system, such
as a combination of localization and spin POVMs (in flat or curved
metric backgrounds).

\item After all these problems have been solved, we'll still have to
find a theory of the quantum dynamics for the spacetime structure.

\end{itemize}

\section*{Acknowledgments and apologies}
We are grateful to numerous friends for helping us locate
references. We apologize if we missed some relevant ones. Only in
a few cases, the omission was intentional.\medskip

Work by AP was supported by the Gerard Swope Fund and the Fund for
Promotion of Research. Part of the work by DRT  was carried at the
Technion---Israel Institute of Technology, and was supported by a
grant from the Technion Graduate School.

{\appendix\section{Relativistic states transformations} In this
Appendix we list the conventions we used and outline the
transformation rules for free particle states.  Details can be
found in the treatises of Bogolubov, Logunov and Todorov (1975),
and Weinberg (1995). Explicit forms of the transformation laws for
massive particles are given by Halpern (1968), Bogolubov
\etal (1975),  and Ahn
\etal\ (2003); for massless particles, by Lindner, Peres, and Terno
(2003) and Bergou, Gingrich, and Adami (2003).

Unless stated otherwise, we chose the following conventions for
states and related operators:
\beq
|\sigma,p\9=\hat{a}^\dag_{\sigma\bp}|\Omega\9,
\eeq
and
\beq
\6\sigma,p|\xi,q\9=(2\pi)^3
(2p^0)\delta_{\sigma\xi}\delta^{(3)}(\bp-\bq),
\eeq
where $p^0\equiv E(\bp)=\sqrt{m^2+\bp^2}$. One-particle states are
\beq
|\Psi\9=\sum_\sigma\int_{-\infty}^\infty
\psi_\sigma(p)|\sigma,p\9d\mu(p), \label{wave}
\eeq
with the Lorentz-invariant measure
\beq
d\mu(p)=d^3\bp/(2\pi)^3(2p^0).
\eeq
The wave functions $|\Psi\9$ satisfy
\beq
\6\sigma,p|\Psi\9=\psi_\sigma(p),\eeq
and
\beq
\6\Psi|\Phi\9=\sum_\sigma\int\psi_\sigma^*(p)\phi_\sigma(p)d\mu(p).
\eeq
If we want to be more explicit about the spin degrees of freedom,
we use 2-spinor notations: a pure state of definite momentum and
arbitrary spin is ${\alpha\choose\beta} |p\9$. The one-to-one
correspondence with Dirac's notation is explained by Bogolubov,
Logunov and Todorov (1975).

Under a classical, geometric Lorentz transformation
$y^\mu=\Lambda^\mu_{\ \nu} x^\nu$, the unitary transformation of
the basis vectors (A.1) is
\beq
U(\Lambda)|\sigma,p\9=\sum_\xi
D_{\xi\sigma}[W(\Lambda,p)]|\xi,\Lambda p\9,
\eeq
where $D_{\xi\sigma}$ are matrix elements of the unitary operator
$D$ that corresponds to the Wigner rotation $W(\Lambda,p)$, given
by Eq.~(\ref{wigwe}) below.

Note that the spin rotation depends on the value of the momentum
(spin is a secondary variable, as defined in Sec.~IV). The quantum
circuit in Fig.~4 gives a graphical representation of primary vs.\
secondary variables.\footnote{This representation was suggested
 to us by Barbara Terhal.}

The Wigner rotation matrix is given by
\beq
W(\Lambda,p):=L^{-1}(\Lambda p)\Lambda L(p), \label{wigwe}
\eeq
where  $L(p)$ is a ``standard boost'' which transforms a
``standard four-momentum'' $k_S$ into $p$. For massive particles
$k_S=(m,0,0,0)$, while for massless ones it is $k_S=(1,0,0,1)$.
Explicit formulas for $L(p)$ in the massive and massless cases are
given in the books of Halpern (1968), Bogolubov
\etal\ (1990), and Weinberg (1995).

Wave functions having a distribution of momenta transform as
\begin{eqnarray}
\nonumber \psi'_\xi(q)&=& \6\xi,q|U(\Lambda)\sum_\sigma\int \!d\mu(p)
\psi_\sigma(p)|\sigma,p\9,\\ \nonumber
&=& \sum_{\sigma,\chi}\int\!
d\mu(p)\psi_\sigma(p)D_{\chi\sigma}[W(\Lambda,p)]\6\xi,q|\chi,\Lambda
p\9,  \\
 &=&\sum_\sigma
D_{\xi\sigma}[W(\Lambda,\Lambda^{-1}p)]\psi_\sigma(\Lambda^{-1}p),
\end{eqnarray}
so the same state in the boosted frame is
\beq
|\Psi'\9=\sum_{\sigma,\xi}\int_{-\infty}^\infty\!
D_{\sigma\xi}[W(\Lambda,\Lambda^{-1}p)]
\psi_\xi(\Lambda^{-1}p)|\sigma,p\9d\mu(p).\label{state}
\eeq
Explicit expressions for $D[W]$ are given in Section IV and in the
references cited above.

\vspace{-5cm}
\begin{figure}[htbp]
\epsfxsize=0.40\textwidth
\centerline{\epsffile{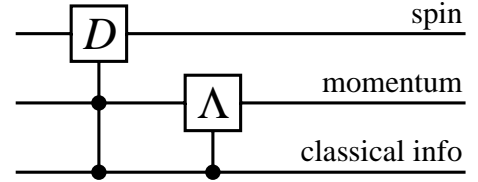}} \vspace*{-5cm} \caption{\small{
Relativistic state transformation as a quantum circuit: the gate
$D$ which represents the matrix $D_{\xi\sigma}[W(\Lambda,p)]$ is
controlled by both the classical information and the momentum $p$,
which is itself subject to the classical information $\Lambda$.}}
\end{figure}

\section{Black hole radiation}

The energy radiated by a black hole satisfies approximately the
Stefan-Boltzmann law, (Frolov and Novikov, 1998; Brout \etal,
1995) so the rate of mass loss due to energy conservation is
\beq
\dot{M}\propto -T^4 A\propto-M^{-2},\label{loss}
\eeq
where $A$ is the  horizon area and time is that of a distant
observer. It can be shown that a relation $T\propto M^{-1}$ holds
in quasi-static changes of mass at all stages of evaporation.
Numerical coefficients were calculated by Page (1976). A back hole
of initial mass $M_0$ (not too small) evaporates after a time
\beq
t_E=aM_0^3,
\eeq
where $a=4.9\times 10^{-9}$sec/kg$^3$. Together with (\ref{loss}),
this gives the following expression for the mass
\beq
M(t)=M_0(1-t/t_E)^{1/3}. \label{mass}
\eeq
The duration of the steady-state radiation build-up is
incomparably shorter than $t_E$ (Wald, 1994; Brout \etal, 1995),
so that the above expression is a good approximation. Hence it
takes a time comparable to the age of the universe for a black
hole of mass $5\times 10^{14}$g (and radius of atomic size) to
evaporate completely (Frolov and Novikov, 1998).

Hawking (1976, 1982) introduced a superoperator (originally called
``super\-scattering operator"), that was mentioned in
Sec.~\ref{terbla} to describe the quantum state evolution during
the black hole formation and evaporation. In standard scattering
theory, a unitary $S$-matrix relates the density matrix of final
states with to of the incoming states:  $\rho^{\rm out}=S\rho^{\rm
in}S^\dag$. For a spacetime with an evaporating black hole,  $S$
would map states from ${\cal H}_{{\rm in}}$ (the states in the
distant past, when the black hole did not exist yet) to the tensor
product of  ${\cal H}_{{\rm out}}$ (the states that reach infinity
and are accessible to a distant observer) and the Hilbert space of
states that fell into the black hole. This splitting is a standard
step in many derivations of the Hawking radiation (Wald, 1994).
Since only the states that reach infinity are accessible to a
distant observer, the final density matrix is  calculated by
tracing out the black hole,
\beq
\rho^{\rm out}=\tr_{\BH} S\rho^{\rm in}S^\dag.
\eeq}

\section*{References}
\small \frenchspacing
\begin{description} \itemsep 0pt

\item Aharonov Y., and D. Z. Albert, 1981, Phys. Rev. D {\bf24}, 359.
\item Aharonov Y., and D. Z. Albert, 1984, Phys. Rev. D {\bf29}, 228.
\item Aharonov, Y., A. Casher, and S. Nussinov, 1987, Phys. Lett. B {\bf
194}, 38.
\item Ahn, D.,  H.-J. Lee, Y. H. Moon, S. W. Hwang, 2003, Phys. Rev. A
{\bf67}, 012103.
\item  Alberghi, G. L., R. Casadio, G. P. Vacca, and G. Venturi, 2001, Phys.
Rev. D {\bf 64}, 104012.
\item Alsing, P. M., and G. J. Milburn, 2002, Quant. Inf.
Comp. {\bf2}, 487.
\item Alsing, P. M., and G. J. Milburn, 2003, e-print
quant-ph/0302179.
\item Amosov, G. G., A. S. Holevo, and R. F. Werner, 2000,
 Probl. Info. Transmission {\bf36}, 305.
\item  Amrein, W. O., 1969, Helv. Phys. Acta {\bf42}, 149.
\item Araki, H., 1999, {\it Mathematical Theory of
 Quantum Fields\/} (Oxford University, Oxford,UK).
\item Ashtekar, A., J. Baez, A. Corichi, and K. Krasnov, 1998, Phys. Rev.
 Lett. {\bf80}, 904.
 \item Audretsch, J., and R. M\"{u}ller, 1994a, Phys. Rev. D {\bf 49},
4056.
\item Audretsch, J., and R. M\"{u}ller, 1994b, Phys. Rev. D {\bf 49},
6566.
\item Bagan, E., M. Baig, and R. Mu\~noz-Tapia, 2001, Phys. Rev. Lett.
{\bf87}, 257903.
\item Balescu, R., and T. Kotera, 1967, Physica (Utrecht) {\bf33}, 558.
\item Ballentine, L. E., 1970, Rev. Mod. Phys. {\bf42}, 358.
\item Bardeen, J. M., B. Carter, and S. W. Hawking, 1973, Comm. Math.
Phys. {\bf31}, 161.
\item Baumgartel, H., and M. Wollenberg, 1992, {\em Causal nets of
operator algebras: mathematical aspects of algebraic quantum field
theory\/} (Akademie, Berlin).
\item Bechmann-Pasquinucci, H., and N.
Gisin, 1999, Phys. Rev. A {\bf59}, 4238.
\item  Bechmann-Pasquinucci, H.,  and A. Peres, 2000, Phys. Rev. Lett. {\bf 85},
3313.
\item  Beckman, D., D. Gottesman, M. A. Nielsen, and J.
Preskill, 2001, Phys. Rev. A {\bf64}, 052309.
\item Beckman, D., D. Gottesman, A. Kitaev, and J.
Preskill, 2002, Phys. Rev. D {\bf65}, 065022.
\item Bekenstein, J. D., 1972, Lett. Nuovo. Cim. {\bf 4}, 737.
\item Bekenstein, J. D., 1974, Phys. Rev. D {\bf9}, 3292.
\item Bekenstein, J. D., 1993, Phys. Rev. Lett {\bf70}, 3680.
\item Bekenstein, J. D., 2002,  in {\it Advances in the Interplay
between Quantum and Gravity Physics\/}, edited by P. G. Bergmann
and V. de Sabbata (Kluwer, Dordrecht, The Netherlands) p.~1
[e-print gr-qc/0107049].
\item Bekenstein, J. D., and A. E. Mayo, 2001, Gen. Rel. Grav. {\bf
33}, 2095.
\item Bell J. S., 1964, Physics {\bf1}, 195.
\item Bell, J. S., and J. M. Leinaas, 1983, Nucl. Phys. B {\bf212}, 131.
\item Bennett, C. H. and G. Brassard, 1984, in {\it Proceedings of IEEE
International Conference on Computers, Systems and Signal
Processing, Bangalore, India\/} (IEEE, New York, 1984) p.~175.
\item Bennett, C. H., G. Brassard, S. Breidbart, and S.
Wiesner, 1983, in {\it Advances in Cryptology\/} (Proceedings of
Crypto-82, Plenum, New York) p.~267.
\item Bennett, C. H., G. Brassard, C. Cr\'{e}peau, R.
Jozsa, A. Peres, and W. Wootters, 1993, Phys. Rev. Lett. {\bf 70},
1895.
\item Bennett, C. H., D. P. DiVincenzo, C. A. Fuchs, T. Mor, E. Rains,
P. W. Shor, J. A. Smolin, and W. K. Wootters, 1999, Phys. Rev. A
{\bf59}, 1070.

\item  Bergou, A. J., R. M. Gingrich, and C. Adami, 2003, e-print
quant-ph/0302095.
\item Bialynicki-Birula, I., 1996, {\it Progress in Optics XXXVI\/},
edited by E. Wolf (Elsevier, Amsterdam), p. 245.
\item Birrell, N. D.,  and P. C. W. Davies, 1982, {\em Quantum Fields
  in Cureved Space} (Cambridge University, Cambridge, UK).
\item Bisognano, J. J., and E. H. Wichmann, 1976, J. Math. Phys.
{\bf17}, 303.
\item Blanchard, Ph., and A. Jadczzik, 1996, Found. Phys. {\bf26}, 1669.
\item Blanchard, Ph., and A. Jadczzik, 1998, Int. J. Theor. Phys.
{\bf37}, 227.
\item Bloch, I., 1967, Phys. Rev. {\bf156}, 1377.
\item Bogolubov, N. N., A. A. Logunov, A. I. Oksak, and I. T. Todorov,
 1990, {\it General Principles of Quantum Field Theory\/} (Kluwer,
Dordrecht, The Netherlands).
\item  Bogolubov N. N., A. A. Logunov, and I. T. Todorov, 1975, {\em
Introduction to Axiomatic Quantum Field Theory} (Benjamin, New
York, NY).
\item Bohm, D., 1951, {\it Quantum Theory}
(Prentice-Hall, New York, NY)
\item Bohr, N., 1927, in {\it Atti del Congresso Internazionale dei
Fisici, Como\/}; reprinted in Nature (London) {\bf121}, 78, 580
(1928).
\item Bohr, N., 1939, in {\it New Theories in Physics,\/} edited by
International Institute of Intellectual Cooperation (Paris).
\item Bohr, N. 1949, in {\it Albert Einstein: Philosopher-Scientist\/},
edited by P. A. Schilpp (Library of Living Philosophers, Evanston,
IL).
\item Bohr, N., and L. Rosenfeld, 1933, Mat. Fys. Medd. Dan. Vidensk.
Selsk. {\bf12} (8).
\item Bombelli, L. R., R. Koul, J. Lee, and R. Sorkin, 1986, Phys. Rev.
 D {\bf34}, 373.
\item  Borde, A., L. H. Ford, and T. A. Roman, 2002, Phys. Rev. D. {\bf
65}, 084002.
\item Boulware, D. G., 1976, Phys. Rev. D {\bf 13}, 2169.
\item Bratteli, O., and D. W. Robinson,  1987, {\em Operator Algebras
and Quantum Statistical Mechanics\/} (Springer, New York, NY) 2
volumes, 2nd edition.
\item Braunstein, S. L., A. Mann, and M. Revzen, 1992, Phys. Rev. Lett.
{\bf68}, 3259.
\item   Brout, R., S. Massar, R. Parentani, Ph. Spindel, 1995,
Phys. Rep. {\bf 260}, 329.
\item  Bru\ss, D., 1998, Phys. Rev. Lett. {\bf81}, 3018.
\item  Bru\ss, D., and Macchiavello, C., 2002, Phys. Rev. Lett. {\bf88},
127901.
\item Buchholz, D.,  and K. Fredenhagen, 1982, Commun. Math. Phys.
{\bf84}, 1.
\item  Busch, P., 1999, J. Phys. A: Math. Gen. {\bf 32}, 6535.
\item Buttler, W. T., R. J. Hughes, S. K. Lamoreaux, G. L. Morgan,
J. E. Nordholt, and C. G. Peterson, 2000, Phys. Rev. Lett.
{\bf84}, 5652.
\item Candelas, P., and D. W. Sciama, 1977, Phys. Rev. Lett. {\bf38},
1372.
\item Casimir, H. G. B., 1948, Proc. Kon. Akad. Wetenschap. {\bf51}, 793.
\item Caves, C. M., 1982, Phys. Rev. D {\bf26}, 1817.
\item  Chuang, I. L, and M. A. Nielsen,  1997, J. Mod. Opt.
{\bf 44}, 2455.
\item Cirel'son, B. S., 1980, Lett. Math. Phys. {\bf4}, 93.
\item Clauser, J. F., M. A. Horne, A. Shimony, and R. A. Holt, 1969, Phys.
Rev. Lett. {\bf23}, 880.
\item Currie, D. G., T. F. Jordan, and E. C. G. Sudarshan, 1963, Rev.
Mod. Phys. {\bf35}, 350.
\item Czachor M., 1997, Phys. Rev. A {\bf55}, 72.
\item Davies, E. B., 1976, {\it Quantum Dynamics of Open
Systems\/} (Academic Press, New York, NY).
\item Davies, P. C. W., 1975, J. Phys. A: Math. Gen. {\bf8}, 609.
\item Davies, P. C. W., T. Dray, and C. A. Manogue, 1996, Phys. Rev. D
{\bf53}, 4382.
\item Detweiler, S., 1982, {\it Black Holes. Selected
Reprints}, (Am. Assoc. of Phys. Teachers, Stony Brook, NY)
\item Dicke, R. H., 1981, Am. J. Phys. {\bf49}, 925.
\item Dieks, D., 1982, Phys. Lett. A {\bf92}, 271.
\item Dirac, P. A. M., 1947, {\it The Principles of Quantum Mechanics\/}
(Oxford University, Oxford, England) p.~36.
\item Drell, S., 1978, Am. J. Phys. {\bf46}, 597; Physics Today {\bf31}
(6), 23.
\item Eggeling, T., D. Schlingemann, and R. F. Werner, 2002,
Europhys. Lett. {\bf57}, 782.
\item Einstein, A., 1949, in {\it Albert Einstein:
Philosopher-Scientist\/}, edited by P. A. Schilpp (Library of
Living Philosophers, Evanston, IL) pp. 85, 683.
\item Einstein A., B.
Podolsky and N. Rosen, 1935, Phys. Rev. {\bf 47}, 777.
\item Eisert J., C. Simon, and M. B. Plenio, 2002, J. Phys. A: Math.
Gen. {\bf 35}, 3911.
\item Emch, G. G., 1972, {\it Algebraic Methods in Statistical Mechanics and
Quantum Field Theory\/} (Wiley-Interscience, New York, NY).
\item Emch, G. G., and C. Liu, 2002, {\it The Logic of Thermostatistical
Physics} (Springer, Berlin).
\item Epstein, H., V. Glaser, and A. Jaffe, 1965, Nuovo Cim. {\bf36},
1016.
\item Finkelstein, D., 1988, in {\it The Universal Turing Machine, A
Half-Century Survey,\/} edited by R. Herken (Oxford University,
Oxford, UK) p.~349.
\item Florig, M., and Summers, S. J., 1997, J. Math. Phys. {\bf38},
1318.
\item Fredenhagen, K., 1985, Commun. Math. Phys. {\bf97}, 461.
\item Frolov, V. P., and I. D. Novikov, 1998, {\it Black Hole Physics\/}
(Kluwer, Dordrecht, The Netherlands).
 \item Frolov, V. P., and D. N. Page, 1993, Phys. Rev. Lett. {\bf71},
3902.
\item Frolov, V. P., and G. A. Vilkovisky, 1981, Phys. Lett. B {\bf 106}, 307.
\item Fuchs, C. A, and A. Peres, 2000, Physics Today {\bf53} (3), 70.
\item Fuchs, C. A., and J. van de Graaf, 1999, IEEE Trans. Info. Theory
\item Fulling, S. A., 1989, {\it Aspects of Quantum Field Theory in
Curved Space-Time\/} (Cambridge University, Cambridge, UK).
\item  Gerlach, U. H., 1976, Phys. Rev. D {\bf 14}, 1479.
\item Giannitrapani, R., 1998, J. Math. Phys. {\bf39}, 5180.
\item  Gingrich, R. M., and C. Adami, 2002, Phys. Rev. Lett.  {\bf 89}, 270402.
\item Gisin, N., G. Ribordy, W. Tittel, and H. Zbinden, 2002, Rev. Mod.
Phys. {\bf74}, 145.
\item Glauber, R. J., 1986, in {\it New Techniques and Ideas in
Quantum Measurement Theory\/}, edited by D. M. Greenberger, Ann.
New York Acad. Sci. {\bf480}, 336.
\item Groisman, B., and B. Reznik, 2002, Phys. Rev. A {\bf66},
022110.
\item Haag, R., 1996, {\it Local Quantum Physics: Fields, Particles,
Algebras\/} (Springer, Berlin).
\item Haag, R., and J. A. Swieca, 1965, Commun. Math. Phys. {\bf1}, 308.
\item Hacyan, S., 2001, Phys. Lett. A {\bf 288}, 59.
\item Halpern, F. R., 1968 {\it Special Relativity and Quantum
Mechanics\/} (Prentice-Hall, Englewood Cliffs, NJ), pp.~80 and
134.
\item Hartle, J. B., 1998, in {\it Black Holes and Relativistic Stars\/},
edited by R. M. Wald (University of Chicago Press, Chicago, IL).
\item Hawking, S. W., 1974, Nature {\bf 248}, 30.
\item Hawking, S. W., 1975, Commun. Math. Phys. {\bf43}, 199.
\item Hawking, S. W., 1976, Phys. Rev. D {\bf14}, 2460.
\item Hawking, S. W., 1982, Commun. Math. Phys. {\bf87}, 395.
\item Hawking, S. W., 1988, Phys. Rev. D {\bf 37}, 904.
\item Hawking, S. W., and G. F. R. Ellis, 1973, {\it The Large Scale
Structure of Space-Time\/} (Cambridge University, Cambridge, UK).
\item Hay, O., and A. Peres, 1998, Phys. Rev. A {\bf58}, 116.
\item Hegerfeldt, G. C., 1985, Phys. Rev. Lett. {\bf 54}, 2395.
\item Herbert, N., 1981, Found. Phys. {\bf12}, 1171.
\item Hillery, M., R. F. O'Connell, M. O. Scully, and E. P. Wigner,
1984, Phys. Rep. {\bf106}, 121.
\item Hod, S., 2002, Phys. Lett. A {\bf299}, 144.
\item Holevo, A. S., 1973,  Probl. Inform. Transmission {\bf9},
110, 177 [transl. from the Russian].
\item Holevo, A. S., 1982,  {\it Probabilistic and Statistical Aspects
of Quantum Theory} (North-Holland, Amsterdam)
\item Holevo, A. S., 1999, Russ. Math. Surveys {\bf53}, 1295.
\item Hosoya A., A. Carlini, and T. Shimomura, 2001, Phys. Rev. D {\bf63},
104008.
\item Ingarden, R. S., 1976, Rep. Math. Phys. {\bf 10}, 43.
\item Ingarden, R. S., A. Kossakowski, and M. Ohaya, 1997, {\it
Information Dynamics and Open Systems\/} (Kluwer, Dordrecht, The
Netherlands).
\item Iorio, A., G. Lambiase, and G. Vitiello, 2001, Ann. Phys. (NY)
 {\bf294}.
\item Jarett, K., and T. Cover, 1981, IEEE Trans. Info. Theory,
{\bf IT-27}, 152.
\item Kemble, E. C., 1937, {\it The Fundamental Principles of Quantum
Mechanics\/} (McGraw-Hill, New York, NY; reprinted by Dover)
p.~244.
\item Kent, A., 1999, Phys. Rev. Lett. {\bf 83}, 1447.
\item Kent, A., 2003, Phys. Rev. Lett. {\bf 90}, 237901.
\item Keyl,  M., 2002, Phys. Rep. {\bf 369}, 431.
\item  Keyl, M., D. Schlingemann, and R. F. Werner, 2003, Quant.
Info. Comp. {\bf 3}, in press [e-print quant-ph/0212014].
\item King, C., and M. B. Ruskai, 2001, IEEE Trans. Info. Theory
{\bf IT-47}, 192.
\item Killing, W. K. J., 1892, J. Reine Angew. Math. {\bf109}, 121.
\item Koopman, B. O., 1931, Proc. Natl. Acad. Sci. U.S.A. {\bf17}, 315.
\item Kraus, K., 1971, Ann. Phys. {\bf 64}, 311.
\item Kraus, K., 1983, {\it States, Effects, and Operations: Fundamental
Notions of Quantum Theory\/} (Springer, Berlin).
\item Landau, L., and R. Peierls, 1931, Z. Phys. {\bf69}, 56.
\item Landau, L. J., 1987, Phys. Lett. A, {\bf120}, 54.
\item Landauer, R., 1991, Physics Today {\bf44} (5), 23.
\item Laplace, P.-S., 1795, {\it Exposition du Syst\`eme du Monde\/},
 (Imprimerie du Cercle-Social, Paris), Vol.~2, p.~305
 [English translation in hawking and Ellis (1973)].
\item Leutwyler, H., 1965, Nuovo Cim. {\bf37}, 556.
\item Levin, O., Y. Peleg, and A. Peres, 1992, J. Phys. A: Math. Gen.
{\bf25}, 6471.
\item Levin, O., Y. Peleg, and A. Peres, 1993, J. Phys. A:
Math. Gen. {\bf26}, 3001.
\item Levitin, L. B., 1969, in {\it Proc.\ Fourth
All-Union Conf.\ on Information and Coding Theory\/}, (Tashkent,
1969) p.~111 [in Russian].
\item Levitin, L. B., 1987, in {\it Information Complexity and Control
in Quantum Physics\/}, ed.\ by A.  Blaqui\`ere, S. Diner, and G.
Lochak (Springer, Vienna, 1987) p.~15.
\item Lindblad, G., 1976,  Commun. Math. Phys. {\bf 48}, 119.
\item Lindner, N. H., A. Peres,
and D.R. Terno, 2003, J. Phys. A {\bf36}, Lxxx [e-print
hep-th/0304017].
\item Lo, H.-K., and H. Chau, 1997, Phys. Rev. Lett. {\bf 78},
3410.
\item  Makeenko, Y., 2002, {\it Methods of Contemporary Gauge Theory}
 (Cambridge University, Cambridge, UK).
\item Mandel, L., 1966, Phys. Rev. {\bf144}, 1071.
\item Mandel, L., and E. Wolf, 1995, {\it Optical Coherence and
Quantum Optics\/} (Cambridge University, Cambridge, UK).
\item Mayers, D., 1997, Phys. Rev. Lett. {\bf 78}, 3414.
\item  Michell, J., 1784, Phil. Trans. R. Soc. (London) {\bf 74},
35 [reprinted in Detweiler (1982)].
\item Mott, N. F., 1929, Proc. Roy. Soc. London A {\bf126}, 79.
\item Neumann, H., and R. Werner, 1983, Int. J. Theor. Phys. {\bf
22}, 781.
\item Newton, T. D., and E. P. Wigner, 1949, Rev. Mod. Phys
{\bf21}, 400.
\item Pachos J., and E. Solano, 2003, Quant. Inf. Comp. {\bf 3}, 115.
\item Parker, S., S. Bose, and M. B. Plenio, 2000, Phys. Rev. A {\bf
61}, 032305.
\item Pechukas, P., 1994, Phys. Rev. Lett. {\bf73}, 1060.
\item Percival, I. C., 1998, Phys. Lett. A {\bf244}, 495.
\item Peres, A., 1980, Phys. Rev. D {\bf22}, 879.
\item Peres A., 1993 {\it Quantum Theory: Concepts and Methods\/}
(Kluwer, Dordrecht, The Netherlands).
\item Peres, A., 1995, in {\it Fundamental Problems in Quantum
Theory\/}, edited by D. M. Greenberger and A. Zeilinger, Ann. New
York Acad. Sci. {\bf755}, 445.
\item Peres A., 1996, Phys. Rev. Lett. {\bf77}, 1413.
\item Peres A., 2000a,  Phys. Rev. A {\bf61}, 022116.
\item Peres A., 2000b,  Phys. Rev. A {\bf61}, 022117.
\item Peres A., 2001,  Phys. Rev. A {\bf64}, 066102.
\item Peres, A., and P. F. Scudo, 2001, Phys. Rev. Lett. {\bf87}, 167901.
\item Peres, A., P. F. Scudo, and D. R. Terno, 2002, Phys. Rev. Lett.
{\bf88}, 230402.
\item  Peres, A., and D. R. Terno, 1998, J. Phys. A: Math. Gen. {\bf 31},
  L671.
\item Peres, A., and D. R. Terno, 2001, Phys. Rev. A {\bf63}, 022101.
\item Peres, A., and D. R. Terno, 2002, J. Mod. Optics {\bf49}, 1255.
\item Peres, A., and D. R. Terno, 2003, J. Mod. Optics {\bf50},
1165.
\item Peres, A., and W. K. Wootters, 1985, Phys. Rev. D {\bf32}, 1968.
\item Peskin, M. E., and D. V. Schroeder, 1995, {\it An Introduction
to Quantum Field Theory\/} (Addison-Wesley, Reading, MA).
\item Preskill, J., 1993, in {\it Black Holes, Membranes, Wormholes and
Superstrings\/},  Superstrings, edited by  S. Kalara and D. V.
Nanopoulos, World Scientific, Singapore) p.~22 [e-print
gr-qc/9209058].
\item Reed, M., and B. Simon, 1980, {\it Functional Analyis},
vol.~1 of {\it Methods of Modern Mathematical Physics} (Academic,
New York, NY).
\item Reeh, H., and S.
Schlieder, 1961, Nuovo Cim. {\bf 22}, 1051.

\item Reznik, B., 2000, e-print quant-ph/0008006.
\item Rivest, R., A. Shamir, and L. Adleman, 1978, Commun. ACM {\bf 21}
(2), 120.
\item Roberts, J. E., 1982, Commun. Math. Phys. {\bf85}, 87.
\item Rosenstein, B., and M. Usher, 1987, Phys. Rev. D {\bf36}, 2381.
\item Salgado D., and J. L. S\'anchez-G\'omez, 2002, e-print
quant-ph/0211164.
\item Scarani, V., W. Tittel, H. Zbinden, and N. Gisin, 2000, Phys.
Lett. A {\bf276}, 1.
\item Schumacher, B., 1995, Phys. Rev. A {\bf51}, 2738.
\item Shannon, C. E., 1948, Bell Syst. Tech. J. {\bf27}, 379, 623.
\item Shor, P., 1994, in {\it Proceedings, 35th Annual Symposium on
Foundations of Computer Science\/} (IEEE, Los Alamitos, CA).
\item  Sonego, S., A. Almergren, M. A. Abramowicz, 2000, Phys.
Rev. D {\bf 62}, 064010.
\item Sorkin, R. D., 1993, in {\it Directions in General Relativity\/},
edited by L. Hu and T. A. Jacobson (Cambridge University,
Cambridge, UK).
\item Spaarnay, M. J., 1958, Physica {\bf24}, 751.
\item \v Stelmachovi\v c, P., and V. Bu\v zek, 2001, Phys. Rev. A
 {\bf64}, 062106.
 \item  Stinespring, W. F., 1955, Proc. Amer. Math. Soc. {\bf 6},
 211.
\item Streater, R. F, and A. S. Wightman, 1964,
{\it PCT, Spin and Statistics, and all that\/} (Benjamin, New
York).
\item Strominger, A., 1996, in {\em Fluctuating Geometries in
Statistical Mechanics and Field Theory, 1994 Les  Houches Summer
School}, edited by F. David, P. Ginsparg, and J. Zinn-Justin
(North-Holland, Amsterdam) [e-print hep-th/9501071].
\item Sudarshan, E. C. G., P. M. Mathews and J. Rau, 1961, Phys.
Rev. {\bf 121}, 920.
\item Summers, S. J., 1990, in {\it Quantum Probability and
Applications V, Lecture Notes in Mathematics 1442\/}, edited by L.
Accardi and W.  von Waldenfels (Springer, Berlin), p. 393.
\item Summers, S. J., and R. Werner, 1985, Phys. Lett. A {\bf
110}, 257.
\item Summers, S. J., and R. Werner, 1987a, J. Math. Phys. {\bf
28}, 2440.
\item Summers, S. J., and R. Werner, 1987b, J. Math. Phys. {\bf
28}, 2447.
\item Terashima, H., and M. Ueda, 2003, Int. J. Quant. Info. {\bf
1},93.
\item Terno, D. R, 2002, in {\it Quantum Theory: Reconsideration of
Foundations\/}, edited by A. Khrennikov (V\"axj\"o University,
V\"axj\"o, Sweden) p.~397. [e-print quant-ph/011144].
\item Terno, D. R., 2003, Phys. Rev. A {\bf 67},
014102.
\item 't Hooft, G., 1996, Int. J. Mod. Phys. A {\bf11}, 4623.
\item 't Hooft, G., 1999, Class. Quant. Grav. {\bf16}, 3263.
\item Tippler, F. G, 1978, Phys. Rev. D {\bf17}, 2521.
\item Toller, M., 1999, Phys. Rev. A {\bf59}, 960.
\item Unruh, W. G., 1976, Phys. Rev. D {\bf14}, 870.
\item Unruh, W. G., and R. M. Wald, 1984, Phys. Rev. D {\bf29}, 1047.
\item Vaidman, L., 2003, Phys. Rev. Lett. {\bf 90}, 010402.
\item  von Borzeszkowski, H., and M. B. Mensky,
2000, Phys. Lett. A {\bf 269}, 197.
\item von Neumann, J., 1932, {\it Mathematische Grundlagen der
Quantenmechanik\/} (Springer, Berlin), p.~236.
\item von Neumann, J., 1955, {\it Mathematical Foundations of Quantum
Mechanics\/}, translated by R. T. Beyer (Princeton University,
Princeton, NJ), p.~418.
\item Wald, R. M., 1975, Comm. Math. Phys. {\bf45}, 9.
\item Wald, R. M., 1984, {\it General Relativity\/} (University of Chicago,
Chicago, IL).
\item Wald, R. M., 1994, {\it Quantum Field Theory in Curved Spacetime
and Black Hole Thermodynamics\/} (University of Chicago, Chicago,
IL).
\item Wald, R. M., 1999, Class. Quant. Grav. {\bf 16}, A177.
\item Wald, R. M., 2001, {\it The Thermodynamics of Black Holes\/}, in
Living Rev. Relativity {\bf4}, 6. [Online article:
http://www.livingreviews.org/Articles/Volume4/2001-6wald].
\item Walgate, J., and L. Hardy, 2002, Phys. Rev. Lett. {\bf89}, 147901.
\item Wehrl, A., 1978, Rev. Mod. Phys. {\bf 50}, 221.
\item Weinberg, S., 1992, {\it Dreams of a Final Theory\/} (Pantheon,
New York, NY).
\item Weinberg, S., 1995, {\it The Quantum Theory of Fields\/}
 (Cambridge University, Cambridge, UK) Vol.~I.

 \item  Werner, R., 1986, J. Math. Phys. {\bf 27}, 793.
 \item Wightman, A. S., 1962, Rev. Mod. Phys. {\bf 34}, 845.
 \item Wiesner, S., 1983, SIGACT News, {\bf 15}, 78.
\item Wigner, E., 1932, Phys. Rev. {\bf40}, 749.
\item Wigner, E., 1939, Ann. Math. {\bf40}, 149.
\item Wootters, W. K., 1998, Phys. Rev. Lett. {\bf80}, 2245.
\item Wootters, W. K., and W. H. Zurek, 1982, Nature {\bf299}, 802.
\item Zbinden, H., J. Brendel, N. Gisin, and W. Tittel, 2000, Phys.
Rev. A {\bf63}, 022111.
\item Zeilinger, A., 1999, Rev. Mod. Phys. {\bf71}, S288.
\item Zel'dovich, Ya. B., 1977, Sov. Phys. JETP {\bf 45}, 9.
\item Zurek, W. H., 1990, editor, {\it Complexity, Entropy, and the
Physics of Information\/} (Santa Fe Institute Studies in the
Sciences of Complexity, vol.~VIII, Addison-Wesley, Reading, MA).
\item Zurek, W. H., 1991, Physics Today {\bf44} (10), 36.
\item Zurek, W. H., 1993, Physics Today {\bf46} (4), 13.
\item Zurek, W. H., 2002, Los Alamos Science {\bf 27}, 2
[e-print quant-ph/0306072].
\item Zurek, W. H., 2003, Rev. Mod. Phys. {\bf75}, 715.

\end{description}

\end{document}